\documentclass[a4paper,11pt]{article}
\pdfoutput=1

\usepackage{jheppub}
\usepackage[latin1]{inputenc}
\usepackage{subcaption,multirow,mathtools,booktabs,xspace}

\graphicspath{{images/}}

\newlength{\rivetLength}
\newcommand{\rivetFigure}[1]{
  \includegraphics[width=\rivetLength]{#1}
}
\newcommand{\rivetFigureRatio}[1]{
  \includegraphics[width=\rivetLength]{#1}\vspace*{-0.5mm}
}

\newcommand{\eqRef}[1]{eq.~\eqref{#1}\xspace}
\newcommand{\eqsRef}[1]{eqs.~\eqref{#1}\xspace}
\newcommand{\EqRef}[1]{Eq.~\eqref{#1}\xspace}

\newcommand{\appRef}[1]{appendix~\ref{#1}\xspace}

\newcommand{\figRef}[1]{figure~\ref{#1}\xspace}
\newcommand{\fig}[1]{figure~#1\xspace}

\newcommand{\FigRef}[1]{Figure~\ref{#1}\xspace}

\newcommand{\secRef}[1]{section~\ref{#1}\xspace}

\newcommand{\SecRef}[1]{Section~\ref{#1}\xspace}

\newcommand{\TabRef}[1]{Table~\ref{#1}\xspace}

\newcommand{\mrm}[1]{\mathrm{#1}}

\renewcommand{\b}{{\mathrm b}}
\renewcommand{\c}{{\mathrm c}}
\renewcommand{\d}{{\mathrm d}}
\newcommand{\e}{{\mathrm e}}

\newcommand{\g}{{\mathrm g}}

\newcommand{\n}{{\mathrm n}}
\newcommand{\p}{{\mathrm p}}
\newcommand{\q}{{\mathrm q}}
\newcommand{\s}{{\mathrm s}}

\renewcommand{\u}{{\mathrm u}}

\newcommand{\D}{{\mathrm D}}

\newcommand{\K}{{\mathrm K}}

\newcommand{\Z}{{\mathrm Z}}

\newcommand{\dbar}{\overline{\mathrm d}}

\newcommand{\pbar}{\overline{\mathrm p}}
\newcommand{\qbar}{\overline{\mathrm q}}

\newcommand{\sbar}{\overline{\mathrm s}}

\newcommand{\ubar}{\overline{\mathrm u}}

\newcommand{\ee}{\e^+\e^-}

\newcommand{\as}{\alpha_{\mathrm{s}}}

\newcommand{\mZ}{m_{\mathrm{Z}}}

\newcommand{\mT}{m_{\perp}}

\newcommand{\pT}{p_{\perp}}

\newcommand{\pTo}{p_{\perp 0}}

\newcommand{\meanPT}{\langle\pT\rangle}
 
\newenvironment{Itemize}{\begin{list}{$\bullet$}%
{\setlength{\topsep}{0.2mm}\setlength{\partopsep}{0.2mm}%
\setlength{\itemsep}{0.2mm}\setlength{\parsep}{0.2mm}}}%
{\end{list}}
\newcounter{enumct}

\title{\boldmath Thermodynamical String Fragmentation}

\author[a,b]{Nadine Fischer}
\author[a]{Torbj\"orn Sj\"ostrand}
\affiliation[a]{Theoretical Particle Physics, 
Department of Astronomy and Theoretical Physics, 
Lund University, SE-223 62 Lund, Sweden}
\affiliation[b]{School of Physics and Astronomy, Monash University, 
Clayton VIC-3800, Australia}
\emailAdd{nadine.fischer@monash.edu}
\emailAdd{torbjorn@thep.lu.se}


\arxivnumber{1610.09818}

\abstract{
The observation of heavy-ion-like behaviour in pp collisions at the LHC
suggests that more physics mechanisms are at play than traditionally
assumed. The introduction e.g.\ of quark-gluon plasma or colour rope
formation can describe several of the observations, but as of yet
there is no established paradigm. In this article we study a few possible
modifications to the \textsc{Pythia} event generator, which describes
a wealth of data but fails for a number of recent observations. Firstly,
we present a new model for generating the transverse momentum of hadrons
during the string fragmentation process, inspired by thermodynamics,
where heavier hadrons naturally are suppressed in rate but obtain a
higher average transverse momentum. Secondly, close-packing of strings
is taken into account by making the temperature or string tension
environment-dependent. Thirdly, a simple model for hadron rescattering
is added. The effect of these modifications is studied, individually and
taken together, and compared with data mainly from the LHC. While some
improvements can be noted, it turns out to be nontrivial to obtain
effects as big as required, and further work is called for.
}

\begin{document}
\setcounter{tocdepth}{2}
\maketitle
\flushbottom

\section{Introduction}

QCD, the theory of strong interactions, is at the origin of a 
wide range of phenomena. In one extreme, progress on high-energy 
perturbative calculations offers an increasingly precise 
and successful description of hard processes, as a large community 
is steadily improving calculational techniques. NLO calculations, 
once rare, are now standard, NNLO is getting there, and even NNNLO 
is starting to appear (see e.g.\ \cite{Salam:2016lhcp} and references 
therein). In another extreme, the nonperturbative aspects of 
low-energy interactions are less well understood.
Lattice QCD can be used to calculate static hadron properties, but not 
(yet?) dynamical processes. Specifically, the description of hadronization, 
the step whereby partons turn into hadrons in high-energy collisions, 
cannot be derived directly from the QCD Lagrangian within any currently 
known formalism. Instead string \cite{Andersson:1983ia} and cluster 
\cite{Field:1982dg,Gottschalk:1983fm,Webber:1983if} models, developed 
in the early eighties, have been used almost unchanged from PETRA/LEP 
$\e^+\e^-$ events to S$\p\pbar$S/Tevatron/LHC $\p\pbar/\p\p$ ones --- 
the assumed "jet universality". Differences have been attributed to the 
quite disparate parton-level configurations that undergo hadronization: 
while $\e^+\e^-$ involves only hard process and final-state radiation
(FSR), pp adds aspects such as initial-state radiation (ISR), 
multiparton interactions (MPIs), beam remnants and 
colour reconnection (CR).   

Cracks have started to appear in this picture as new LHC data have been
presented. Specifically, several studies have shown how high-multiplicity
pp events have properties similar to those observed in heavy-ion AA
collisions. Some observations may have an explanation within the current 
framework, e.g.\ CR may give some flow-like patterns \cite{Ortiz:2013yxa}, 
but others do not.
An early example was the discovery of ``the ridge'', an 
enhanced particle production around the azimuthal angle of a trigger 
jet, stretching away in (pseudo)rapidity \cite{Khachatryan:2010gv,%
Aad:2015gqa,Khachatryan:2016txc}. A more recent example
is the smoothly increasing fraction of strange baryon production with
increasing charged multiplicity, a trend that lines up with pA data 
before levelling out at the AA results \cite{Adam:2016emw}
\footnote{Note that in \cite{Cuautle:2016huw} the authors obtain 
the same enhancement regardless of the system measured if the 
yield ratios are plotted against the estimated energy density.}. 
Conventional wisdom holds that the formation of a quark--gluon plasma 
(QGP) requires a larger volume and longer time for thermalization than 
pp or pA systems can offer, so such trends are unexpected, see
e.g.~\cite{BraunMunzinger:2007zz}. 

It is therefore time to rethink the picture of hadronization in 
high-energy and high-multiplicity collisions. One possible approach 
is to imagine that a QGP is at least partly formed in pp collisions,
such that individual colour fields (strings) cease to exist. Such a 
behaviour is already implemented in the EPOS model \cite{Pierog:2013ria}. 
Another is to imagine that strings survive as a vehicle e.g. of 
short-range flavour correlations, but that their properties are modified. 
Colour ropes \cite{Biro:1984cf,Bialas:1984ye,Andersson:1991er} is one 
such example, wherein several colour-triplet strings combine to a higher 
colour-representation field. A detailed implementation of rope dynamics 
is found in the DIPSY program \cite{Bierlich:2014xba}. Both EPOS and DIPSY 
qualitatively describe several of the new key features, such as the 
increasing rate of strangeness production at higher multiplicities.
 
With the studies described in this article we want to add to the set
of alternative models that can be used to compare with data. At best
it may offer some new insights, at worst it will act as a straw man
model. Firstly, rather than the particle-mass-independent Gaussian 
$\pT$ spectrum assumed in the standard string model, it introduces an 
exponential $\pT$ dependence, $\exp(-\pT / T)$. This is split among 
possible flavours according to hadronic $\mT$, $\exp(-\mT / T)$. 
Such $\pT$ and $\mT$ shapes were used to describe early pp data,
e.g.\ at the ISR \cite{Capiluppi:1973fz}, and has some foundation
in the Hagedorn temperature
\cite{Hagedorn:1965st,Hagedorn:1970gh,Hagedorn:1983wk} and in
related \cite{Barshay:1972jy} ideas. (Later powerlike $\pT$ ans\"atze
\cite{Hagedorn:1983wk, Tsallis:1987eu,Cleymans:2016opp} or two-component
exponential + powerlike ones \cite{Bylinkin:2012bz} can be viewed
as a consequence of perturbative jet production, and is in our
framework generated as such, in an earlier stage than the
nonperturbative hadronization.) Secondly, it assumes that the close-packing 
of several strings leads to an increased effective temperature and 
thereby both a changed particle composition and changed $\pT$ spectra. 
In spirit this is close to the rope model, but it does not have to assume 
that the individual strings either fuse or melt away. Thirdly, if the 
fragmenting strings are close-packed this also implies the initial
formation of a dense hadronic gas, wherein rescattering may lead to 
collective-flow effects. Such effects are simulated in a crude first 
approximation.

The impact of these mechanisms on experimental distributions is studied,
in order to quantify their significance. As a prime example, consider the
$\meanPT (n_{\mathrm{ch}})$ distribution, with a characteristic
rising trend that has been proposed as a signal for colour reconnection
\cite{Sjostrand:1987su}. Alternative interpretations are now offered
in terms of close-packing of strings and/or hadrons, and these are 
presented and compared with data individually. At the end of the day,
we should expect the ``true'' nature of high-multiplicity $\p\p$ 
collisions to contain many contributing mechanisms, however. To be more
more specific, in quantum mechanics any process that is not explicitly
forbidden by some selection rule is bound to occur, the question is only 
with what rate. The final task therefore is try to constrain the relative 
importance of the mechanisms, not to prove a specific one 
``right'' or ``wrong''.

The new model components are implemented as options in the standard 
\textsc{Pythia} event generator \cite{Sjostrand:2006za,Sjostrand:2014zea}, 
which makes it easily accessible for further experimental tests.
They should be viewed as a first iteration. Should they prove useful 
there is room for further improvements, as we will indicate.

The article is organized as follows. \SecRef{sec:existModels} outlines relevant 
features of the existing Lund string model and introduces key 
observables, with emphasis on those new ones that are not well 
described by the current \textsc{Pythia} generator. \SecRef{sec:newModels}
introduces the alternative approaches explored in this article, 
and presents some first toy studies for simplified string topologies. 
Comparisons with data are presented in \secRef{sec:results}, highlighting 
what seems to work where and what not. Finally \secRef{sec:summary} contains 
a summary and outlook. 

\section{Existing Models and Data}
\label{sec:existModels}

\subsection{The Lund string model}

The Lund string fragmentation model \cite{Andersson:1983ia} is very 
successful in many respects, but more so for the overall longitudinal 
fragmentation structure than for its description of the particle 
composition.

The central assumption in the string model is that of linear confinement,
$V(r) = \kappa r$, with a string tension $\kappa \approx 1$~GeV/fm.
The word ``string'' should here not be taken literally; the physical
object is a kind of flux tube stretched between the endpoints, with
a typical transverse size of the order of the proton one, 
$r_{\p} \sim 0.7$~fm. The one-dimensional ``mathematical'' string should 
then be viewed  as a description of the location of the center of the 
flux tube. By analogy with superconductivity the tube could be viewed as 
a vortex line like in a type II superconductor, alternatively as an 
elongated bag in a type I one.

In the case of a simple stable back-to-back $\q\qbar$ system, with 
$m_{\q} = p_{\perp\,\q} = 0$, quarks move with the speed of light in 
``yo-yo''-mode oscillations, as energy moves between being stored
in the endpoint quarks and in the intermediate string. If creation 
of new $\q\qbar$ pairs is allowed the original system can break up into
smaller ones, each a colour singlet in its own right. Denoting the 
original pair $\q_0\qbar_0$, and ordering the new pairs 
$\q_i\qbar_i, 1 \leq i \leq n - 1$ from the quark end, results 
in the production of $n$ hadrons $\q_0\qbar_1$, $\q_1\qbar_2$, \ldots, 
$\q_{n-1}\qbar_0$.

Aligning the $x$ axis with the string axis, the breakup vertices are
characterized by their location $(t_i, x_i)$. These vertices have a 
spacelike separation, and so have no unique time ordering. (Except for 
the original $(t_0, x_0) = (0, 0)$ of course. But here it is actually 
the turning points of the $\q_0$ and $\qbar_0$ that define the vertices
in \eqRef{eq:lundvtxsep} below, and then spacelike separation is 
restored.) Two adjacent ones are correlated by the constraint that the 
hadron produced should have the correct mass $m_i$:
\begin{equation}
\kappa^2 ((x_i - x_{i-1})^2 - (t_i - t_{i-1})^2) = m_i^2 ~.
\label{eq:lundvtxsep}
\end{equation}
If the vertices are assigned from the quark end, say, each new vertex
therefore corresponds to one degree of freedom, which should be selected
according to some probability function. Imposing consistency 
constraints, mainly that results should be the same (on the average) 
if fragmentation is instead considered from the antiquark end, gives
the solution~\cite{Andersson:1983ia}
\begin{equation}
f(z) \propto \frac{1}{z} \, (1-z)^a \, \exp(-bm^2 / z) ~,
\label{eq:lundFragFun}
\end{equation}
with $a$ and $b$ two free parameters, and where $m^2 \to \mT^2$ once
transverse momentum is introduced. Here $z$ is the fraction of 
available lightcone momentum $E + p_x$ taken by a hadron, with the
remainder $1 - z$ retained by the string for subsequent particle
production. 

This ansatz leads to vertices having an equilibrium distribution 
(after having taken a few steps away from the endpoints)
\begin{equation}
P(\Gamma) \propto \Gamma^a \, \exp(-b\Gamma) ~,~~~~~~~ 
\Gamma = (\kappa \tau)^2 = \kappa^2 (t^2 - x^2)  ~,
\end{equation}
with  the same $a$ and $b$ as above. (For the special case $a = 0$ 
this result agrees with the Artru-Mennessier model \cite{Artru:1974hr},
which is based on constant decay probability per string area $\d t \, \d x$,  
without any mass constraint.) The associated probability for producing 
$n$ particles can be written as \cite{Andersson:1985qr}
\begin{equation}
\d P_n \propto \left[ \prod_{i = 1}^n N \d^2 p_i \delta(p_i^2 - m_i^2) \right]
\, \delta^{(2)} \left( \sum_i p_i - p_{\mathrm{tot}} \right) \,
\exp(- b \kappa^2 A_{\mathrm{tot}}) ~,
\end{equation}
where $A_{\mathrm{tot}}$ is the total space--time area under the 
breakup vertices. The relation between $\d P_n$ and $\d P_{n-1}$ 
(at a reduced c.m. energy) is then given by the fragmentation function 
\eqRef{eq:lundFragFun}, where it is easy to show that the exponentials 
match, and somewhat less trivial that a larger $N$ (i.e. larger weight for 
higher multiplicities) corresponds to a larger $a$ (i.e. less momentum 
taken away in each step).

The simple $\q\qbar$ fragmentation picture can be extended to $\q\qbar\g$
topologies if the gluon is viewed as having separate colour and anticolour 
indices, as in the $N_C \to \infty$ limit \cite{'tHooft:1973jz}. Then one
string piece is stretched between the quark and the gluon, and another 
between the gluon and the antiquark. The absence of a string piece stretched 
directly between the quark and antiquark leads to predicted asymmetries in the 
particle production \cite{Andersson:1980vk} that rapidly were observed 
experimentally \cite{Bartel:1981kh}. In general, a string can stretch 
from a quark end via a number of intermediate gluons to an antiquark end.
Technically the motion and fragmentation of such a string system can become 
rather complicated \cite{Sjostrand:1984ic}, but the fragmentation can be 
described without the introduction of any new principles or parameters.
This is the most powerful and beautiful aspect of the string fragmentation 
framework. Note that the leading hadron in a gluon jet can take momentum 
from both the string pieces that attaches it to colour-adjacent partons. 
This is unlike cluster models, where gluons are forced to branch into 
$\q\qbar$ pairs, such that smaller colour singlets are formed rather than
one single string winding its way between the partons. The string model 
is easily extended to closed gluon loops and, with rather more effort 
\cite{Sjostrand:2002ip}, to junction topologies, where three string pieces 
come together in a single vertex.

We now turn to the breakup mechanism. If a $\q\qbar$ pair is massless 
and has no transverse momentum it can be produced on-shell, in a single 
vertex, and then the $\q$ and $\qbar$ can move apart, splitting the string 
into two in the process. But if the $\q$ (and $\qbar$) transverse mass 
$m_{\perp \q} = \sqrt{m_{\q}^2 + p_{\perp \q}^2} > 0$ this is no longer 
possible. By local flavour conservation the $\q\qbar$ pair is still 
produced at a common vertex, but as virtual particles that each needs 
to tunnel out a distance $d = \mT / \kappa$. Using the WKB 
approximation~\cite{Andersson:1983ia} to calculate the tunneling 
probability for the pair gives a factor
\begin{equation}
\exp \left( - \pi m_{\perp \q}^2 / \kappa \right) 
= \exp \left( - \pi m_{\q}^2 / \kappa \right) \,
\exp \left( - \pi p_{\perp \q}^2 / \kappa \right) ~,
\label{eq:tunneling}
\end{equation}
where the Gaussian answer allows a convenient separation of the $m$ and
$\pT$ dependencies (with implicit phase space $\d^2\pT$). 

The latter is implemented by giving the $\q$ and $\qbar$ opposite and 
compensating $\pT$ kicks, with $\langle p_{\perp \q}^2 \rangle = \kappa / \pi 
= \sigma^2 \approx (0.25~\mathrm{GeV})^2$. A hadron receives its $\pT$ 
as the vector sum of it $\q$ and $\qbar$ constituent kicks, and thus 
$\langle p^2_{\perp\mathrm{had}} \rangle = 2  \sigma^2$. 
Empirically the tuned $\sigma$ value comes out larger than this, actually
closer to $\sigma = 0.35$~GeV. This implies that almost half of the 
$\pT^2$ kick is coming from other sources than tunneling. One source 
could be soft gluon radiation below the perturbative (parton shower) 
cutoff, where $\as$ becomes so big that perturbation theory breaks down
\cite{Andersson:1981xu}. Effectively radiation near the 
perturbative/nonperturbative border is thus shoved into an artificially 
enhanced tunneling answer, with the further assumption that the Gaussian 
shape and the $\pT$ balancing inside each new $\q\qbar$ pair still holds. 

Uncertainties also arise in the interpretation of the mass suppression
factor of \eqRef{eq:tunneling}: what quark masses to use? If current 
quark masses then the $\u$ and $\d$ ones are negligible while the $\s$ 
is below 0.2~GeV, predicting less strangeness suppression than observed, 
while with constituent masses $m_{\u} \approx m_{\d} \approx 0.33$~GeV 
and $m_{\s} \approx 0.51$~GeV \cite{Close:1979bt} too much suppression 
is predicted. Intermediate masses and suppression factors closer to data 
can be motivated e.g. by noting that an expanding string corresponds to 
confinement in the two transverse dimensions but not in the longitudinal 
one. In the end, however, the $\s/\u$ suppression is viewed as an empirical
number to be tuned to data. Whichever values are used, $\c$ and $\b$ quark
tunneling production is strongly suppressed, so this  mechanism can be 
totally neglected relative to the perturbative ones.

Considering only mesons in radial and rotational ground states, i.e.\
only the pseudoscalar and vector multiplets, naive spin counting predicts
relative rates $1 : 3$, whereas data prefers values closer to $1 : 1$, 
at least for $\pi : \rho$. It is possible to explain a suppression 
of the vector mesons based on the difference in the hadronic wave 
functions, from the spin--spin interaction term \cite{Andersson:1983ia},
but the amount has to be tuned to data. And further brute-force suppression
factors are needed specifically for the $\eta$ and $\eta'$ mesons, which
have ``unnaturally'' large masses owing to the $U(1)$ anomaly.  
 
Baryon production can be introduced by allowing diquark--antidiquark 
breakups of the string \cite{Andersson:1981ce}, to be viewed as occurring 
in two consecutive $\q\qbar$ creation steps \cite{Andersson:1984af}. 
A baryon and the matching antibaryon would normally be nearest neighbours
along the string, but the ``popcorn mechanism'' also allows one (or more)
mesons to be produced in between. Diquark masses can be used to derive 
approximate suppressions, but again free parameters are used, for 
$\q\q / \q$, $\s\q / \q\q$, $\q\q_1 / \q\q_0$ and others. Unfortunately 
the tuned values do not always match so well with the tunneling-formula 
expectations.

In total $\mathcal{O}(20)$ parameters are used to describe the outcome
of the string/tunneling mechanism for particle production. Notable is 
that the particle masses do not enter explicitly in these considerations.
This is unlike cluster models, e.g., where hadron masses occur in the phase 
space available for different cluster decay channels. A fair overall 
description of the particle composition is then obtained with very few 
parameters \cite{Platzer:2011bc,Richardson:2012bn}. Note that while most 
fragmentation parameters in \textsc{Herwig++} exist in different copies 
for light $(\u,\d,\s)$, $\c$, and $\b$ quarks, the ones for heavy quarks 
have either been set equal to the values of those for light quarks 
\cite{Platzer:2011bc} or have not been included in further tuning processes 
\cite{Richardson:2012bn}.

The hadron masses can be explicitly introduced into the Lund framework 
by assuming that the integral $\int_0^1 f(z) \, \d z$, with $f(z)$ given
by \eqRef{eq:lundFragFun}, provides the relative normalization of 
possible particle states. This concept has been developed successfully 
within the UCLA model \cite{Chun:1997bh,Abachi:2006qf}, in that particle
rates come out quite reasonably with minimal further assumptions. There
are some other issues with this approach, however, and we do not pursue
it further here.

\subsection{Key data}
\label{sec:keydata}

An immense number of studies have been published based on hadron collider 
data, and it is not the intention here to survey all of that. Instead
we here bring up some of the key data and distributions that have 
prompted us to this study. Several of them will be shown repeatedly in the 
following. We note that all histograms we will present in this article are 
produced by utilizing \textsc{Rivet}~\cite{Buckley:2010ar}.

The list of key observables includes:
\begin{itemize} 

\item The change of flavour composition with event multiplicity. 
Specifically, high-multiplicity events have a higher fraction of 
heavier particles, meaning particles with a higher strangeness content 
\cite{Adam:2016emw}. \textsc{Pythia} contains no mechanism to generate 
such a behaviour. On the contrary, within a single fixed-energy string 
a higher multiplicity means more lighter particles, for phase space
reasons. In $\p\p$ collisions a higher multiplicity is predominantly 
obtained by more MPIs, however, so the composition stays rather constant.

\item The average transverse momentum $\meanPT$ is larger
for heavier particles, both at RHIC~\cite{Abelev:2008ab} and 
LHC~\cite{Abelev:2014qqa}. This is a behaviour that is present also 
in \textsc{Pythia}, and comes about quite naturally e.g. by lighter
particles more often being decay products, with characteristic
$\meanPT$ values smaller than the primary particles in the 
string fragmentation. The mass dependence is underestimated, however.
That is, $\pi^{\pm}$ obtains a too large $\meanPT$ in
\textsc{Pythia} and baryons a too small one. Recently $\meanPT$ 
has also been presented as a function of $n_{\mathrm{ch}}$, 
inclusive~\cite{Aad:2010ac} and for different hadron 
species~\cite{Bianchi:2016szl}, 
providing a more differential information on this mismatch.
In \figRef{fig:pTvsMorNch} we show these observables and compare default
\textsc{Pythia} with data, with the above expected conclusions.
Note that the data in~\fig{1} of~\cite{Bianchi:2016szl} is not (yet) publicly
available. To obtain an estimate of the data that is comparable to MC
predictions we used an estimate of the logarithmic fits shown 
in~\fig{1} of~\cite{Bianchi:2016szl} and used $n_{\mathrm{ch}}$ values on the
$x$ axis rather than $\langle\d n_{\mathrm{ch}}/\d\eta\rangle_{|\eta|<0.5}$.

\begin{figure}[p!]
\centering
\rivetFigure{data/ATLAS_2010_S8918562/d25-x01-y01-pTnCh-PyDef.pdf} \hfill
\rivetFigure{data/ALICE_2014_I1300380/d03-x01-y01-pTm-PyDef.pdf} \\[2mm]
\begin{subfigure}[b]{0.49\textwidth}
  \rivetFigureRatio{data/ALICE_2016/d02-x01-y08-pTnCh-Omega-PyDef.pdf}
  \rivetFigureRatio{data/ALICE_2016/d02-x01-y07-pTnCh-Xi-PyDef.pdf}
  \rivetFigureRatio{data/ALICE_2016/d02-x01-y06-pTnCh-Lambda-PyDef.pdf}
  \rivetFigureRatio{data/ALICE_2016/d02-x01-y05-pTnCh-phi-PyDef.pdf}
  \rivetFigureRatio{data/ALICE_2016/d02-x01-y04-pTnCh-ks-PyDef.pdf}
  \rivetFigureRatio{data/ALICE_2016/d02-x01-y03-pTnCh-p-PyDef.pdf}
  \rivetFigureRatio{data/ALICE_2016/d02-x01-y02-pTnCh-k-PyDef.pdf}
  \rivetFigure{data/ALICE_2016/d02-x01-y01-pTnCh-pi-PyDef.pdf}
\end{subfigure}
\hfill
\begin{subfigure}[b]{0.49\textwidth}
  \rivetFigureRatio{data/ALICE_2016/d02-x01-y08-pTnCh-Omega-PyDef-ref.pdf}
  \rivetFigureRatio{data/ALICE_2016/d02-x01-y07-pTnCh-Xi-PyDef-ref.pdf}
  \rivetFigureRatio{data/ALICE_2016/d02-x01-y06-pTnCh-Lambda-PyDef-ref.pdf}
  \rivetFigureRatio{data/ALICE_2016/d02-x01-y05-pTnCh-phi-PyDef-ref.pdf}
  \rivetFigureRatio{data/ALICE_2016/d02-x01-y04-pTnCh-ks-PyDef-ref.pdf}
  \rivetFigureRatio{data/ALICE_2016/d02-x01-y03-pTnCh-p-PyDef-ref.pdf}
  \rivetFigureRatio{data/ALICE_2016/d02-x01-y02-pTnCh-k-PyDef-ref.pdf}
  \rivetFigure{data/ALICE_2016/d02-x01-y01-pTnCh-pi-PyDef-ref.pdf}
\end{subfigure}
\\
\caption{The mean transverse momentum as a function of the charged 
multiplicity (\textit{top left} and the hadron mass (\textit{top right}) 
and \textit{bottom}). Predictions of default \textsc{Pythia} compared to 
ALICE~\cite{Bianchi:2016szl,Abelev:2014qqa} and ATLAS~\cite{Aad:2010ac} 
data. The data in the bottom plots is taken to be an estimate of the 
logarithmic fits in \cite{Bianchi:2016szl} and therefore no error bars
are included.
\label{fig:pTvsMorNch}}
\end{figure}

\item The charged particle $\pT$ spectrum is not correctly modelled 
at low $\pT$ scales, with \textsc{Pythia} producing too few particles 
at very low values \cite{Aad:2010ac,Chatrchyan:2011av,Adam:2015pza}. 
Often tunes then compensate by producing a bit too many at intermediate 
$\pT$ scales. The issue shows up e.g.\ in minimum-bias 
$\d n_{\mathrm{ch}} / \d \eta$ distributions, where it is not possible to 
obtain a good description for data analyzed with $\pT > 0.1$~GeV and 
$\pT > 0.5$~GeV simultaneously.

\item In the $\pT$ spectra for identified particles \cite{Adam:2015qaa}
it turns out that the deficit at low $\pT$ is from too little $\pi^{\pm}$
production. This is not unexpected, given the previous two points,
but stresses the need to revise the mass dependence of $\pT$ spectra.

\item The $\Lambda/\K$ $\pT$ spectrum ratio, measured by
CMS~\cite{Khachatryan:2011tm}, where \textsc{Pythia} is not able to 
reproduce the peak at $\sim 2.5$~GeV completely 
and overshoots the distribution for large-$\pT$ values.

\item The observation of a ridge in $\p\p$ collisions was one of the 
major surprises in the 7~TeV data \cite{Khachatryan:2010gv}, and has
been reconfirmed in the 13~TeV one \cite{Aad:2015gqa,Khachatryan:2016txc}.
The ridge is most clearly visible at the very highest multiplicities,
but more careful analyses hints the effect is there, to a smaller extent, 
also at lower multiplicities. Like in heavy-ion collisions one may
also seek a description in terms of correlation functions,
$C(\Delta\phi) \propto 1 + \sum_{n \geq 2} v_n \, \cos(n\Delta\phi)$,
notably the $v_2$ coefficient, with a similar message. These phenomena 
are not at all described by \textsc{Pythia}: there is no mechanism that 
produces a ridge and, once the effects of back-to-back jet production 
have been subtracted, also no rise of $v_2$.
\end{itemize}

There are also some other reference distributions that have to be 
checked. These are ones that already are reasonably well described,
but that inevitably would be affected by the introduction of new 
mechanisms.
\begin{itemize}

\item The charged particle multiplicity distribution $P(n_{\mathrm{ch}})$
is sensitive to all mechanisms in minimum-bias physics, but especially 
the MPI and CR modelling. A mismatch in $\langle n_{\mathrm{ch}} \rangle$
is most easily compensated by modifying the $\pTo$ scale of the MPI
description. This parameter is used to tame the $\d \pT^2 / \pT^4$ 
divergence of the QCD cross section to a finite 
$\d \pT^2 / (\pTo^2 + \pT^2)^2$ shape. It can 
be viewed as the the inverse of the typical colour screening distance
inside the proton. A mismatch in the width of the $n_{\mathrm{ch}}$
distribution can be compensated by a modified shape of the $b$ 
impact-parameter distribution of the two colliding protons. 
Specifically, a distribution more sharply peaked at $b = 0$ gives a 
longer tail towards high multiplicities.

\item An $\meanPT$ increasing with $n_{\mathrm{ch}}$ was 
noted already by UA1 \cite{Albajar:1989an}, and has remained at 
higher energies \cite{Aad:2010ac,Bianchi:2016szl}. It offers a key 
argument for introducing CR 
in $\p\p / \p\pbar$ collisions, as follows \cite{Sjostrand:1987su}. 
The tail towards large $n_{\mathrm{ch}}$ is driven by events with more
MPI activity, rather than e.g.\ by events with higher-$\pT$ jets.
If each MPI subcollision produces particles essentially independently
the $\meanPT(n_{\mathrm{ch}})$ would be rather flat. 
CR implies that fewer and fewer extra particles are produced for 
each further MPI, as the possibilities to reduce the total string 
length by CR increase the more partons are already present. 
The amount of $\pT$ from the MPIs thus increases 
faster than the $n_{\mathrm{ch}}$, meaning more $\pT$ per particle.
(To this comes the normal hadronization $\pT$ contribution, which
raises the overall $\meanPT$ level but does not contribute
to the $\meanPT(n_{\mathrm{ch}})$ slope.) The exact nature 
of CR is not known, meaning that many models have been developed
\cite{Sjostrand:1987su,Sjostrand:1993hi,Christiansen:2015yqa}.
In most of them there is some overall CR strength parameter that can 
be adjusted to fit the $\meanPT(n_{\mathrm{ch}})$ slope.

\item A natural reference for hadronization properties always is
$\e^+\e^-$ data. The principle of jet universality --- or, in our case,
string universality --- is deeply rooted, so it it useful to check 
that no changes of fundamental string properties have too adverse
an impact on $\e^+\e^-$. There is also a possibility of improvements 
in some places, like the inclusive $p_{\perp\mathrm{in}}$ and 
$p_{\perp\mathrm{out}}$ spectra; unfortunately these are not available
for identified particles. 
\end{itemize}

\section{The New Models}
\label{sec:newModels}

In this section we outline the basic ideas and implementations 
that we have developed to offer new options to the traditional 
\textsc{Pythia} hadronization framework. As we later compare with data 
we will have reason to go into more detail and discuss some variations. 

\subsection{Variations of  the normal string model}
\label{sec:oldModelVar}

As described above, the standard tunneling framework suggests a 
Gaussian suppression of the production of heavier quarks and diquarks, 
with a further suppression based on the hadronic spin state, but no obvious
room for an explicit dependence on the hadron mass. It also 
provides a common Gaussian $\pT$ spectrum for all new $\q\qbar$ pairs.
We will study a few variations of this framework, mainly as a
reference for the thermodynamical ansatz below.
 
Firstly, consider a Gaussian suppression associated with the masses 
of the produced hadrons rather than with the quarks. That is, let the 
relative production rate of different hadron species be given by a factor 
$\exp(-m^2_{\perp\mathrm{had}}/2\sigma^2)$, which factorizes into a 
species-independent $\pT$ spectrum and an 
$\exp(-m_{\mathrm{had}}^2 / 2\sigma^2)$ mass suppression. The question is 
then whether this would give the appropriate suppression for the production 
of heavier particles.

Secondly, the universal $\pT$ spectrum could be broken by assigning 
a larger width in string breakups of the $\s\sbar$ and $\q\q\qbar\qbar$
kinds, relative to the baseline $\u\ubar$ and $\d\dbar$ ones. The issue
to understand here is how dramatic differences are required to get a 
better description of the individual $\pi$, $\K$ and $\p$ $\pT$ 
spectra.

Thirdly, assume that more MPIs leads to a closer packing of strings
in the event, but that each string ``flux tube'' remains as a separate
entity. The transverse region of the string shrinks and, 
essentially by Heisenberg's uncertainty relations, this should 
correspond to a higher energy, i.e. a larger string tension $\kappa$.   
(Such a relation comes out naturally e.g. for bag models of 
confinement \cite{Chodos:1974je}.) Overall the dense-packing effect on 
$\kappa$ and related parameters should scale as some power of 
$n_{\mathrm{MPI}}$, i.e. the number of MPIs in the current event. Since 
$n_{\mathrm{ch}}$ and $n_{\mathrm{MPI}}$ are strongly correlated it is 
thus interesting to study how the particle composition and 
$\meanPT$ depend on $n_{\mathrm{ch}}$.
For a more differential picture it should be preferable to estimate the 
number of strings in the neighbourhood of each new hadron being produced. 

This is done by making a reasonable guess for the momentum of the hadron
that is the next to be produced on the current string. 
Using an average hadron mass and $\pT$, defined in the frame of the parent 
string, and an average $\Gamma$ value of $\langle\Gamma\rangle=(1+a)/b$,
the momentum of the ``average expected'' hadron is calculated. Using this 
information, we determine the number of strings that cross the rapidity of 
the expected hadron. For this purpose the rapidity range that a string
will populate is defined by the rapidity of the endpoint partons of each 
string piece,
\begin{equation}
y=\mrm{sgn}\left(p_z\right)\log\frac{E+|p_z|}
{\sqrt{\mrm{max}\left(m_\perp^2,m_\mrm{min}^2\right)}}~,
\end{equation}
where $m_\mrm{min}^2$ has the purpose to protect against strings with 
low-$m_\perp$ endpoints from populating the full rapidity range.
The rapidity-density measure is reasonable for low-$\pT$ hadroproduction,
but does not reflect the phase space inside a high-$\pT$ jet, where
close-packing of strings should be rare. Therefore the effective number of 
strings is calculated as
\begin{equation}
n_{\mathrm{string}}^{\mathrm{eff}}=1+\frac{n_{\mathrm{string}}-1}
{1+p^2_{\perp\mathrm{had}}/p^2_{\perp\,0}}~,
\label{eq:nStringEff}
\end{equation}
where $p_{\perp\mathrm{had}}$ is the physical hadron $\pT$ and 
$p_{\perp\,0}$ is the MPI regularization parameter.

\begin{figure}[t!]
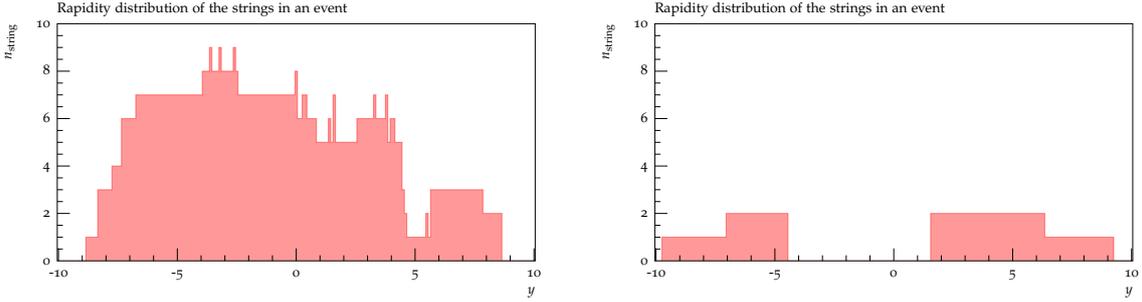

\centering
\rivetFigure{oldModelVariations/stringQCD.pdf} \hfill
\rivetFigure{oldModelVariations/stringDiff.pdf} \\
\caption{Rapidity distribution of the strings (added on top of each other)
in a typical QCD event (\textit{left}) and in a diffractive event 
(\textit{right}).
\label{fig:stringRapDist}}
\end{figure}

As two examples, the rapidity distribution of the strings in a typical QCD 
event and in a diffractive event are shown in \figRef{fig:stringRapDist}.
Using \eqRef{eq:nStringEff}, the string tension in \eqRef{eq:tunneling}
is modified to be
\begin{equation}
\kappa~\to~\left(n_{\mathrm{string}}^{\mathrm{eff}}\right)^{2r}
~\kappa~,
\label{eq:nStringEffKappa}
\end{equation}
where the exponent $r$ is a left as a free parameter, that can be used 
to tune the model to data.
Note that while junctions\footnote{A junction topology corresponds to an Y 
arrangement of strings, i.e. where three string pieces have to be joined up 
in a common vertex.} contribute to the calculation of $n_{\mathrm{string}}$ 
by assuming one string stretched between the highest- and lowest-rapidity 
parton, their fragmentation does not make use of \eqRef{eq:nStringEffKappa}.
Junctions are rare in the models we study, so this is not a significant
simplification.

The effect of modifying the string tension due to the local density has
also been studied in other Monte Carlo programs, which are primarily
for heavy-ion collisions. Some of them have hardly been used for pp physics 
as they miss out on other physics aspects such as QCD jet production.
In the RQMD model~\cite{Sorge:1989dy} for studying relativistic 
nucleus-nucleus collisions, colour strings are allowed to fuse into 
ropes if they are overlapping, which weakens the suppression of strangeness 
and baryon production due to the increased string tension~\cite{Sorge:1992ej}.
A similar model with string fusion into colour ropes is implemented in
the DIPSY event generator~\cite{Bierlich:2014xba} and shows improvement
in the description of identified particle spectra in pp minimum bias data.
For the UrQMD model~\cite{Bass:1998ca,Bleicher:1999xi}, used for 
relativistic heavy-ion and hadron-hadron collisions, the authors 
of~\cite{Soff:2000ae} show that a better description of particle yields is 
achieved with an enhanced string tension. The effect of an increased string 
tension in a densely populated environment on strangeness and diquark 
production, antibaryon-to-baryon ratios and other observables has been 
investigated in~\cite{Soff:2002bn}. In AMPT~\cite{Zhang:1999bd,Lin:2004en}, 
a Monte Carlo transport model for heavy-ion collisions at relativistic 
energies, parameters in the Lund string fragmentation model have been
modified, as the string tension is expected to be increased in the 
dense matter formed in heavy-ion collisions~\cite{Lin:2000cx}.
In the PSM~\cite{Amelin:1993cs,Amelin:2001sk} Monte Carlo model for 
simulating nuclear collisions, string fusion associated with high string 
densities is taken into account to reduce multiplicities and increase 
$\langle p_\perp \rangle$, baryon and strangeness production.
Ref.~\cite{Braun:1992ss} presents a model which introduces the interaction 
between strings via their fusion and percolation analytically. The 
$\langle p_\perp \rangle$ of the produced particles, and therefore also 
the string tension, depends on the string density and how much strings 
overlap~\cite{Braun:1999hv,Braun:2000hd,Pajares:2005kk}.

\subsubsection{One-string toy model}

\begin{figure}[t!]
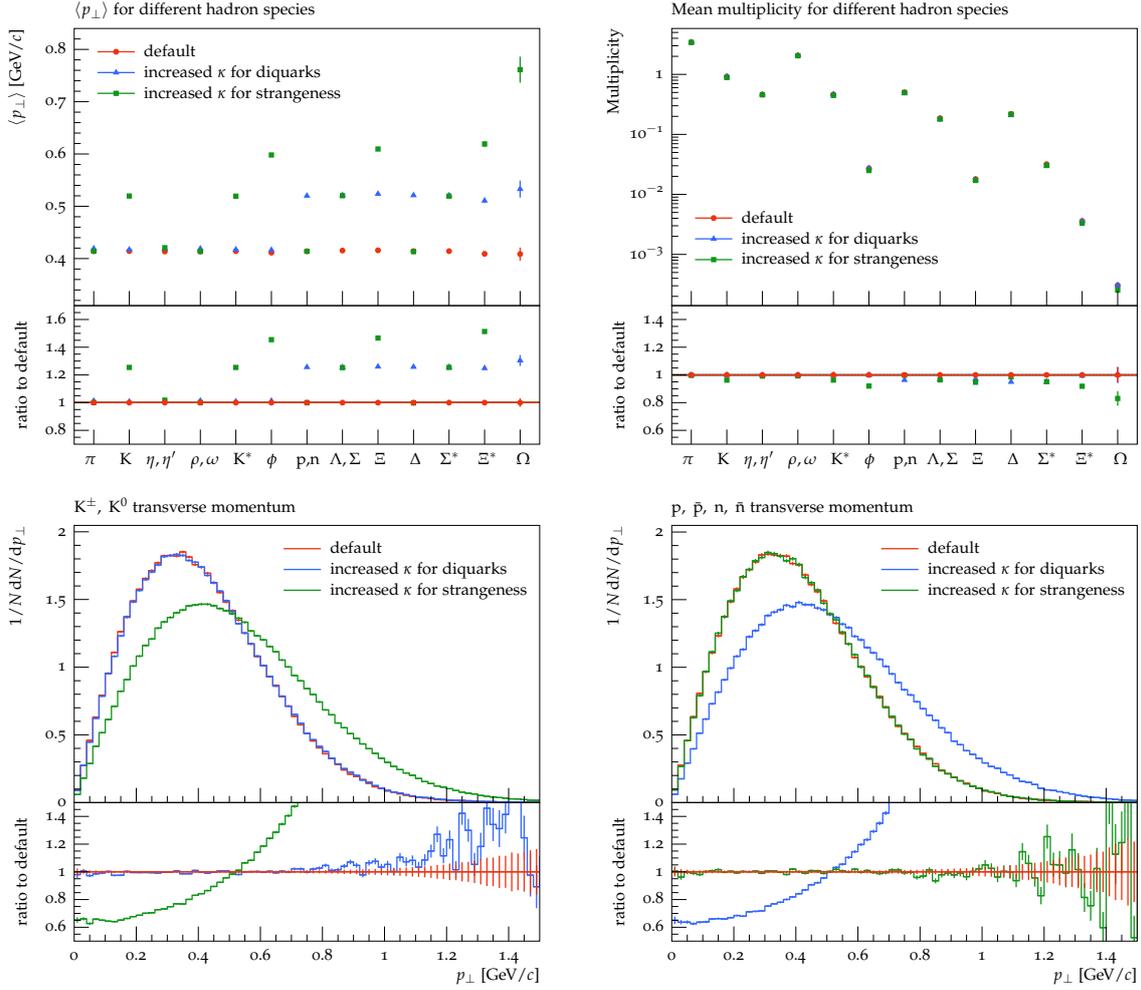

\centering
\rivetFigure{oldModelVariations/avPTtoyOldVarKincrease.pdf} \hfill
\rivetFigure{oldModelVariations/meanMultiToyOldVarKincrease.pdf} \\
\rivetFigure{oldModelVariations/pTkToyOldVarKincrease.pdf} \hfill
\rivetFigure{oldModelVariations/pTpnToyOldVarKincrease.pdf} \\
\caption{$\meanPT$ (\textit{top left}) and mean multiplicity
(\textit{top right}) for different hadron species and the $\K$ 
(\textit{bottom left}) and $\p/\n$ $\pT$ (\textit{bottom right}) spectra 
in the toy model. Predictions of the conventional
string model without modifications are shown in red and with the string
tension $\kappa$ increased for diquarks in blue and strangeness in green.
\label{fig:PTtoyVar1}}
\end{figure}

A very simple toy model is introduced to validate the modifications to the 
string tension in the conventional string model. A single string 
with energy $\mZ$ is spanned along the $z$ axis. The flavour of the 
endpoint quarks is chosen random from the set $(\u,\d,\s,\c,\b)$.
The study includes only primary produced hadrons, i.e. no hadron decays, 
and also excludes the hadrons containing the endpoint quarks.
(Such hadrons would have lower $\meanPT$ since the endpoint quarks
by definition have $\pT=0$.)

The $\meanPT$ and the mean multiplicity for different hadron 
species are shown in \figRef{fig:PTtoyVar1}. As expected, increasing the 
string tension either for $\s$ quarks or for diquarks leads to an 
increased $\meanPT$ value for the hadrons concerned. Note that for 
$\eta+\eta'$ the $\meanPT$ is only increased slightly due to the 
$\u\ubar+\d\dbar$ quark component being more frequently produced compared to 
$\s\sbar$. There is a slight reduction of the production probability for 
hadrons with $\s$ quarks or diquarks, shown in the top 
right plot in \figRef{fig:PTtoyVar1}, due to the increased string tension 
leading to fewer particles being produced in affected events. The bottom 
row of \figRef{fig:PTtoyVar1} shows the $\K$ and $\p/\n$ $\pT$ spectra, 
shifted to larger values as the string tension for that hadron species is
increased.

\subsubsection{Multi-string toy model}

\begin{figure}[t!]
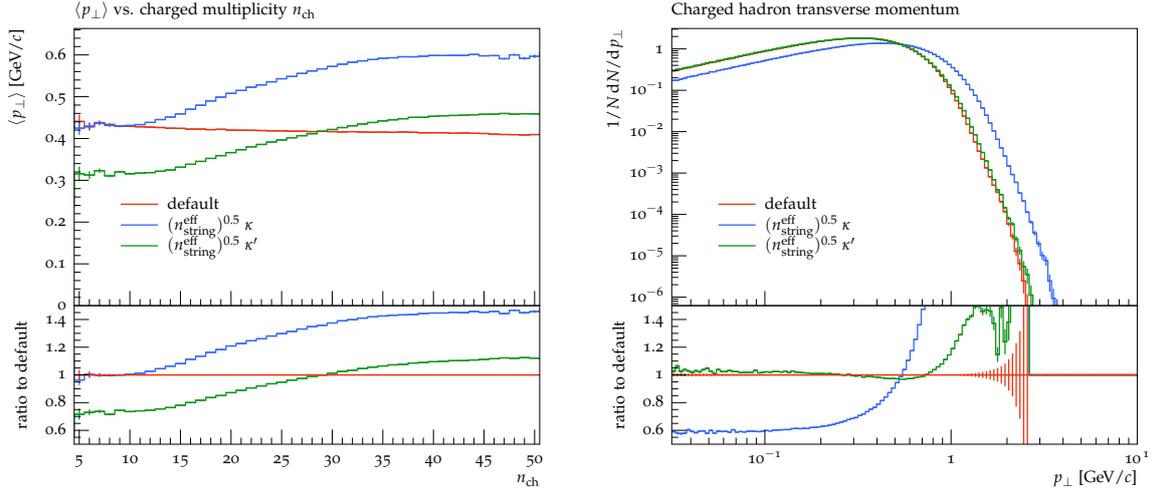

\centering
\rivetFigure{oldModelVariations/avPTvsNchToyOldNst.pdf} \hfill
\rivetFigure{oldModelVariations/ptToyOldNst.pdf} \\
\caption{$\meanPT$ as a function of the number of charged particles
(\textit{left}) and the $\pT$ distribution (\textit{right}) for the toy
model with multiple strings along $z$ axis. Predictions of the default model 
are shown in red and dependence of the string tension on the number of close 
strings in blue and green with two different string tensions $\kappa>\kappa'$.
\label{fig:PTtoyVar2}}
\end{figure}

To investigate the effect of the close-packing of strings, as in 
\eqRef{eq:nStringEffKappa}, the above toy model is extended to include
several strings along the $z$ axis. The number of strings is picked randomly
between two and eight and the string energies are chosen such that they
sum up to $1~\mathrm{TeV}$. \FigRef{fig:PTtoyVar2} shows $\meanPT$ 
as a function of the number of charged particles and the $\pT$ distribution 
and compares the modified model to default \textsc{Pythia}. Two different 
choices for the baseline value for the string tension are made in case of 
taking the close-packing of strings into account. In the first case the 
tension is denoted with $\kappa$ and its value is adjusted such that 
$\meanPT$ agrees with default \textsc{Pythia} for small values of 
$n_\mathrm{ch}$. In the second case, where the string tension is denoted by 
$\kappa'$ the value is adjusted to obtain the same $\meanPT$, 
averaged over all hadrons and charged multiplicities. The latter case serves 
as a cross check when investigating the influence on the $\pT$ spectrum of 
charged hadrons.

As expected the $\meanPT$ increases with the 
charged multiplicity, eventually flattening out at large multiplicities.
The left histogram in \figRef{fig:PTtoyVar2} also nicely shows that
the rise is independent of the baseline string tension value.

When fitting the string tension such that the same overall $\meanPT$
is reached as in the default model, the charged hadron $\pT$ spectrum
exhibits only small changes; making the spectrum somewhat broader.

\subsubsection{Gaussian $m^2_{\perp\mathrm{had}}$ suppression}

\begin{figure}[t!]
\centering
\includegraphics[width=1.45\rivetLength]
{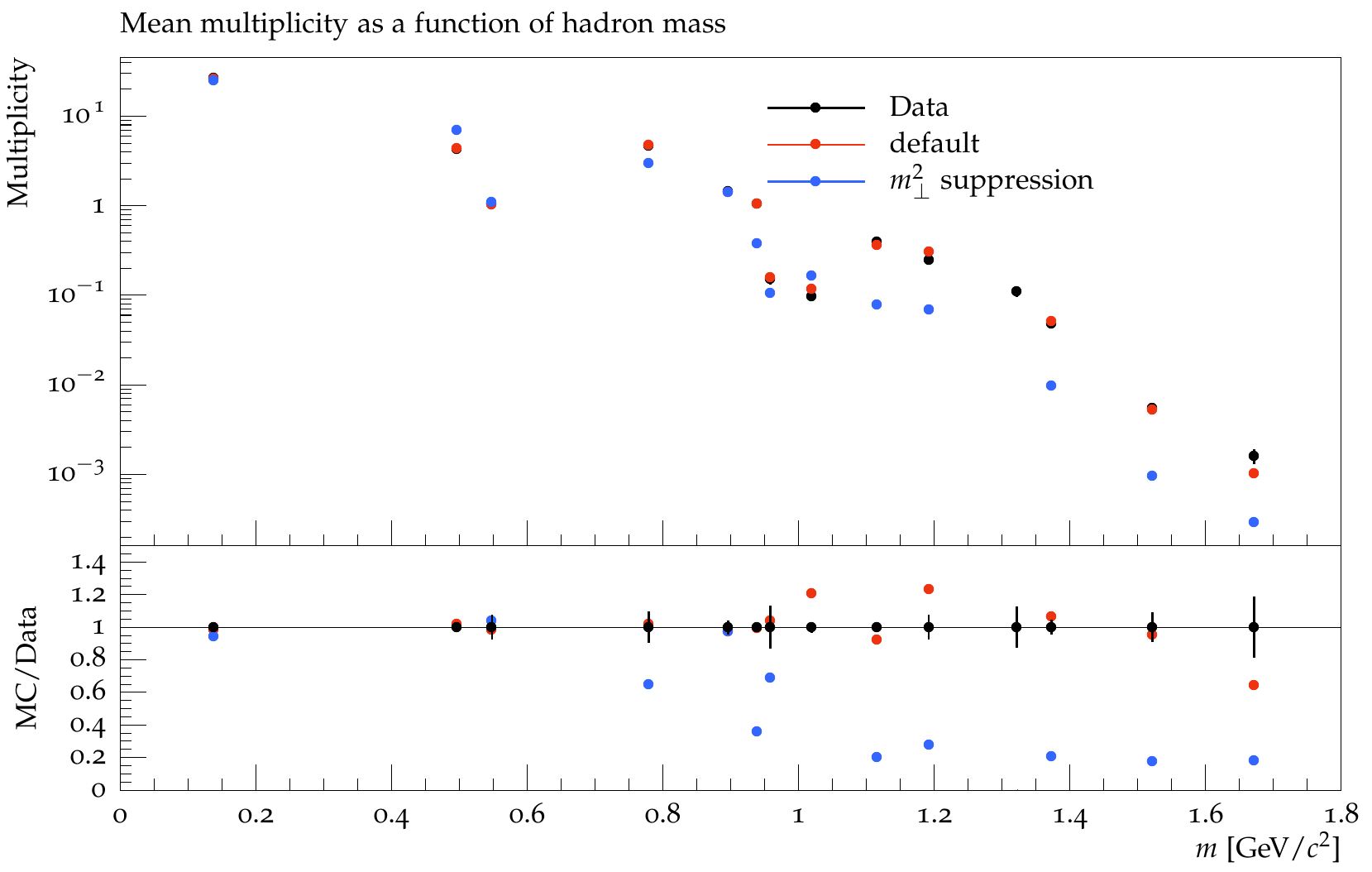} \\
\caption{The conventional string model with its default options
(red) and with the relative production rate of different hadron species 
given by a factor $\exp(-m^2_{\perp\mathrm{had}}/2\sigma)$ (blue),
compared to PDG data~\cite{Amsler:2008zzb}.
\label{fig:meanMultiMT2}}
\end{figure}

To test the applicability of the Gaussian transverse mass suppression,
the quark $\pT$ is generated according to $\exp\left(-p_{\perp \q}^2/
\sigma^2\right)$, see \eqRef{eq:tunneling}, with the hadron flavour chosen
based on $\exp\left(-m^2_{\perp\mathrm{had}}/2\sigma^2\right)$. The additional 
factor of two arises from the hadron receiving $\pT$ contributions from 
two quarks. As the comparison to data is of interest here, realistic 
$\ee\to\mathrm{jets}$ events with $s=\mZ^2$ are investigated. 
In \figRef{fig:meanMultiMT2} the particle composition is shown as a function 
of mass. 
This clearly indicates that the suppression based on the transverse mass 
squared of the hadrons is suppressing heavier hadrons too much. We will 
therefore not consider this option further.

\subsection{The thermodynamical string model}
\label{sec:thermModel}

The most radical departure from standard Lund string principles 
that we explore in this article is to replace the Gaussian suppression 
factor in mass and $\pT$ by an exponential one. To be more explicit,
instead of a quark-level suppression governed by \eqRef{eq:tunneling}
there will be a hadron-level suppression 
\begin{equation}
\exp( - m_{\perp\mathrm{had}} / T ) ~~~~~ \mathrm{with} ~~
m_{\perp\mathrm{had}} = \sqrt{ m_{\mathrm{had}}^2 + p_{\perp\mathrm{had}}^2  }~.
\end{equation}
The inspiration clearly comes from a thermodynamical point of view,
which is why we choose to associate the dimensional parameter with a 
temperature $T$. This association should not be taken too literally,
however; there are many differences relative to a purely thermal model.
The main one is that we keep the longitudinal string fragmentation 
structure unchanged, which ensures local flavour conservation.
Another is that e.g.\ the Hagedorn approach 
\cite{Hagedorn:1965st,Hagedorn:1970gh} is based on the assumption of 
a steeply increasing density of excited states as a function of mass,
whereas we only include a few of the lowest multiplets. (By default
only the ground states corresponding to no radial or orbital excitation,
optionally also the lowest $L = 1$ meson multiplets.) This means that, 
although our $T$ comes out to be a number of the order of the Hagedorn 
temperature, there is no exact correspondence between the two.
Also, $T \sim \sqrt{\kappa / \pi} = \sigma$ from dimensional 
considerations, so our $T$ could be viewed as a manifestation of the
string energy per unit length, not directly linked to a temperature. 

There is also an experimental historical background to the choice of 
an exponential shape, in that already fixed-target and ISR data showed 
that a distribution like $\exp(-B \pT)$ offered a good fit to the inclusive 
$\d n_{\mathrm{ch}} / \d\pT^2$ spectrum, with $B \approx 6~\mathrm{GeV}^{-1}$
\cite{Capiluppi:1973fz,Alper:1975jm,Guettler:1976fc,Giacomelli:1979nu}. 
With data split by particle type, a lower $B$ value is noted for kaons and 
protons than for pions, but with the modified form $\exp(-B \mT)$ all the 
spectra  can be described by almost the same $B \approx 6$ value.

As an aside, the preference for an exponential shape was and is not a 
show-stopper for the Gaussian approach in the normal string fragmentation. 
At larger $\pT$ the spectrum is dominated by the fragmentation of (mini)jets, 
giving a larger rate than the nonperturbative hadronization one. And 
at smaller $\pT$ the pattern of decays makes the spectrum more steep 
than the Gaussian one of the primary hadrons. So at the end of the day 
a Gaussian ansatz lands not that far away from an exponential spectrum,
although differences remain. See further \secRef{sec:effectDecays}, in 
particular \figRef{fig:hadPTdecays}.
 
In more detail, our model is intended to give each new hadron in the
string fragmentation a $\pT$ according to an exponential distribution.
We want to preserve the concept of local $\pT$ conservation in each
$\q\qbar$ breakup vertex, so seek a distribution that convoluted with
itself (in two transverse dimensions) gives an exponential,
\begin{equation}
f_\mathrm{had}\left(p_{\perp\mathrm{had}}\right) =
\exp\left(-p_{\perp\mathrm{had}}/T\right) =
\int\d^2p_{\perp\,1} f_\q(p_{\perp\,1})\int\d^2p_{\perp\,2}f_\q(p_{\perp\,2})
~\delta({\mathbf{p}_{\perp\mathrm{had}}}-{\mathbf{p}_{\perp\,1}}-
{\mathbf{p}_{\perp\,2}})~.
\end{equation}
Using Fourier transforms to turn the convolution into a product,
\begin{align}
\tilde{f}_\mathrm{had}(b_\perp)
&= \frac1{2\pi}\int f_\mathrm{had}\left(p_{\perp\mathrm{had}}\right)
\exp\left(-i\mathbf{b}_\perp\cdot{\mathbf{p}_{\perp\mathrm{had}}}\right)
\d^2p_{\perp\mathrm{had}} \nonumber \\
= 2\pi\,\tilde{f}_\q^2(b_\perp)
&= \frac{1}{\left(1+(b_\perp T)^2\right)^{3/2}}~.
\end{align}
The transformation back of $\tilde{f}_\q$ then gives~\cite{Gradshteyn:1980}
\begin{equation}
f_\q(p_{\perp\,\q}) \propto \int
\frac{\exp(i b (p_{\perp\,\q}/T) \cos\varphi)}{(1 + b^2)^{3/4}}\,
b\,\d b\,\d\varphi \propto
\int_0^{\infty} \frac{b\,J_0(b\,p_{\perp\,\q}/T)} {(1 + b^2)^{3/4}}\, \d b
\propto \frac{K_{1/4}(p_{\perp\,\q}/T)}{(p_{\perp\,\q}/T)^{1/4}} ~,
\label{eq:bessel}
\end{equation}
where $b=b_\perp T$, $J_0$ is a regular Bessel function of the first kind,
and $K_{1/4}$ is the modified Bessel function of the second kind of order
$1/4$. An implementation of $K_{1/4}$ has been included in \textsc{Pythia}
based on~\cite{Abramowitz:1965}, using a power series for $p_{\perp\,\q}/T < 2.5$
and an asymptotic expansion for $p_{\perp\,\q}/T  > 2.5$.

Consider the fragmentation of a string, where the quark $\q$ of one breakup 
has a certain $\mathbf p_{\perp\,1}$. The transverse momentum 
$\mathbf p_{\perp\,2}$ of the (di)quark of the next breakup pair 
$\q'\bar\q'$ is constructed by picking its absolute value according to 
\eqRef{eq:bessel} and a random azimuthal angle. The partner anti(di)quark 
must thus have $-\mathbf p_{\perp\,2}$ due to local momentum conservation.
The hadron transverse momentum is simply the sum of the $\pT$ of the
two contributing quarks, $\mathbf p_{\perp\mathrm{had}}=\mathbf 
p_{\perp\,1}-\mathbf p_{\perp\,2}$. Having $p_{\perp\mathrm{had}}$ at hand 
we decide on the flavour of the breakup pair $\q'\bar\q'$, and therefore 
also on the hadron species, as follows: calculate the transverse mass 
$m_{\perp\mathrm{had}}$ of all hadrons whose flavour content includes the 
incoming quark $\q$ and determine the basic probability for each hadron as
\begin{equation}
P_\mathrm{had}=\exp(-m_{\perp\mathrm{had}}/T)~.
\label{eq:flavPickProb}
\end{equation}
Assuming the production of two hadrons with different masses $m_1$ and $m_2$,
then \eqRef{eq:flavPickProb} implies the same production rate for 
$\pT \gg m_1,m_2$, but more suppression of the heavier hadron at low $\pT$. 
Thus there is less production of heavier states, but they come with a larger
$\meanPT$.

As mentioned above, by default we only include the light-flavour $(\u,\d,\s)$ 
meson and baryon multiplets without radial or orbital excitation
\footnote{Heavy flavour hadrons are of course included to handle the endpoint
quarks of the strings, where needed.}.
However, if desired, more hadrons can be added to the procedure.
Depending on the flavour content of the hadron, the probability in
\eqRef{eq:flavPickProb} receives additional multiplicative factors:
\begin{Itemize}
\item Due to spin-counting arguments vector mesons receive a factor of 
$3$ and tensor mesons a factor of $5$.
\item For same-flavour mesons we include the diagonal meson mixing factors, 
similar to what has been done previously in the conventional Lund string model.
\item Baryons receive a free overall normalization factor with respect to 
mesons, as well as an additional factor stemming from the $SU(6)$ symmetry
factors, see \cite{Andersson:1981ce}. The relative weight of spin $1/2$ 
baryons with respect to those with spin $3/2$ is $2:4$, similar to the factors 
for mesons arising from the spin-counting arguments in point 1.
\item For the special case of octet baryons with three different flavours, 
e.g. $\Lambda$ and $\Sigma^0$, their probability for different internal spin
configurations is taken into account.
\item An extra suppression factor for hadrons with strange (di)quarks is 
included to get more control over the relative hadron production and thus
a better description of data.
\end{Itemize}
All probabilities are then rescaled to sum up to unity and the hadron species
and therefore the flavour of the next (di)quark pair is chosen accordingly.
Note that we have not (yet) implemented popcorn baryon production, i.e.
no mesons are produced in between a baryon and its antibaryon partner.

Similar to \eqRef{eq:nStringEffKappa} the temperature can be modified as
\begin{equation}
T~\to~\left(n_{\mathrm{string}}^{\mathrm{eff}}\right)^r
~T~,
\label{eq:nStringEffTemp}
\end{equation}
with $n_{\mathrm{string}}^{\mathrm{eff}}$ given in \eqRef{eq:nStringEff}
to take into account the effect of close-packed strings.
Note that in~\cite{DiasdeDeus:2006xk} the temperature has been related 
to the density in the context of the percolation of color sources
(the density is however defined differently).

\subsubsection{Asymmetry in different flavour transitions}

\begin{figure}[tbp]
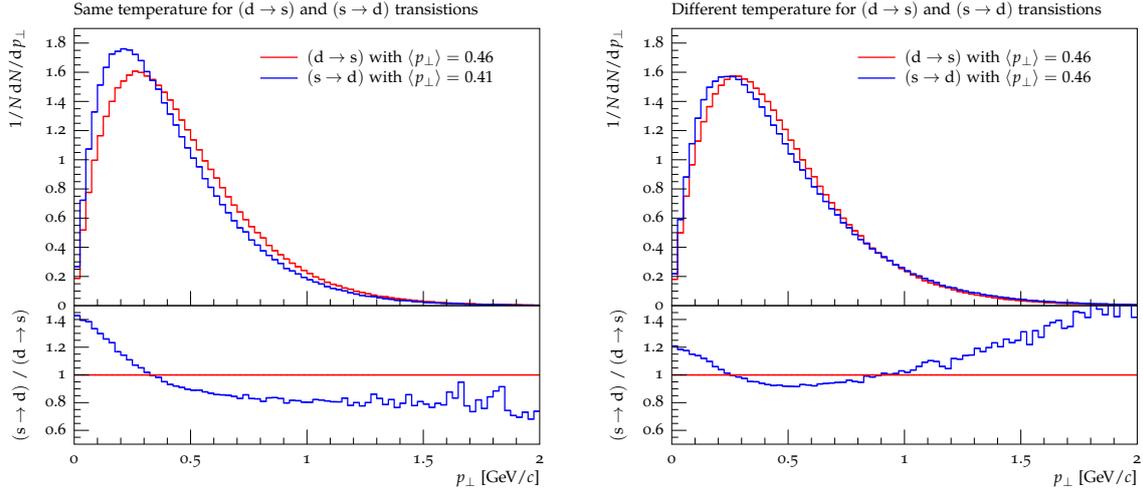

\centering
\rivetFigure{thermalFlavTransAsymm/pT-DtoS-StoD-diffPt.pdf} \hfill
\rivetFigure{thermalFlavTransAsymm/pT-DtoS-StoD-samePt.pdf} \\
\caption{The inclusive hadron $\pT$ spectrum for $(\d\to\s)$ and $(\s\to\d)$
transitions with the same temperature for both cases (\textit{left})
and with the adjusted temperature (\textit{right}).
\label{fig:flavTrans}}
\end{figure}

Consider a very simple model, where only string breaks with $\d\dbar$ and 
$\s\sbar$ quark pairs are allowed to produce only pseudoscalar mesons, and 
the mixing of diagonal mesons is ignored. Then it is rather easy to see 
that the $\pT$ spectra and $\meanPT$ of the hadrons produced in $(\d\to\s)$ 
transitions is not the same as for $(\s\to\d)$ transitions, due to the 
difference in competition. In the first instance $(\d\to\s)$ competes with 
$(\d\to\d)$, and since the former produces the heavier meson it also 
obtains the higher $\meanPT$. In the latter instance $(\s\to\d)$ instead
competes with $(\s\to\s)$ and so gives the lighter meson and lower $\meanPT$.
Assuming that fragmentation is performed from the quark end inwards, 
$\K^0=\d\sbar$ would thus obtain a harder $\pT$ spectrum than $\bar\K^0=\dbar
\s$, which should not be the case. A simple solution for obtaining the same 
$\meanPT$ for both $(\d\to\s)$ and $(\s\to\d)$ transitions is to adjust the 
temperature in \eqRef{eq:flavPickProb} in case of initial $\s/\bar\s$ quarks 
such that $\d\sbar$ and $\dbar\s$ hadrons are produced with the same $\meanPT$ 
value, higher than $\d\dbar$, and $\s\sbar$ becomes even higher than that.

In \figRef{fig:flavTrans} we show the $\pT$ spectra for both types of
transitions with the same temperature and with the adjusted temperature.
Note that though both transitions end up with the same $\meanPT$ 
value, the shape of the $\pT$ distribution still differs somewhat.

Unfortunately this is a price to pay for working with a recursive model,
where flavour is conserved locally. A traditional thermal model based on
\eqRef{eq:flavPickProb} would not conserve flavour or momentum, however,
so is not an option here.

\subsection{The hadronic rescattering model}
\label{sec:hadronScattering}

A close-packing of fragmenting strings also implies a close-packing of
the produced primary hadrons, i.e. a dense hadronic gas. This gives the
possibility for hadrons to rescatter on the way out, in particular at
the earliest times after hadronization. A detailed simulation of this
mechanism would require a knowledge of where in space--time each hadron
is produced. For a single string, say stretched along the $z$ axis, it 
is straightforward to translate between the $(E, p_z)$ values of the 
primary hadrons and the $(t, z)$ coordinates of the string breakups.
For the more realistic case, when a string is stretched between 
several partons and the string motion is considerably more complicated
\cite{Sjostrand:1984ic}, appropriate rules have not been worked out.
To this should be added ambiguities in the transverse production 
coordinates, both as a consequence of the transverse distribution of 
the MPIs and of transverse fluctuations inside each string. The modelling 
of all of these aspects is an interesting task for the future.
In addition, the cross section for the scattering of two hadrons
against each other varies between hadron kinds, and depends on the
relative energy of the two, adding a further layer of complexity. 

Here we want to avoid such a detailed model, but still be able to 
explore whether hadronic rescattering effects could contribute to 
the resolution of some of the effects that we are attempting to explain. 
Collective flow --- whether dictated by properties of the QGP 
or by hadronic rescattering --- is well-established in heavy-ion 
collisions, see e.g. \cite{Gyulassy:2004zy,Heinz:2013th,Roland:2014jsa} 
and references therein. In particular, a common average radial velocity 
means that heavier particles have  a higher $\meanPT$ than 
lighter. Of course we do not expect as dramatic effects in $\p\p$,
but they may still contribute to the same kind of $\pi/\K/\p$ $\pT$
separation as in the thermodynamic scenario above, so it should be
interesting to compare the two possibilities. 

The simple modelling we have in mind is applied to the primary hadrons
produced directly from the string fragmentation, before secondary decays
are considered. Furthermore, for strings stretched along the $z$ axis
there is a strong correlation between the rapidity $y$ of a particle 
and its space--time production vertex. Therefore, for a given hadron,
the density of other hadrons at around the same rapidity is a reasonable 
(and longitudinally boost invariant) measure of how close-packed particle 
production is. If there is a contribution from particles coming from the
same simple string it has presumably already been absorbed in the tuned 
fragmentation parameters, so we should disregard such pairs. Unlike
$\e^+\e^-$ events, however, it is common with topologies where a string
consists of pieces stretched back and forth across the same rapidity 
range, and then the above argument does not apply. In practice, it is
therefore more relevant to exclude rescattering only between close 
neighbours in the fragmentation chains. 

One should further note that the rapidity density of hadrons refers to 
low-$\pT$ particles. The hadronization of a scattered high-$\pT$ parton
mainly occurs at larger $\pT$ scales, and these hadrons would be 
essentially unaffected. 

The angular distribution of a rescattering, defined in the rest frame
of the hadronic pair, should depend on the orbital angular momentum $L$.
For simplicity, we restrict to $s$-wave isotropic scattering (i.e., $L=0$) 
by requiring that the classical value of angular momentum 
$L = b \, |\mathbf{p}| <  b \, |\mathbf{p}_\mrm{max}|\sim 1$, where $b$ 
is the impact parameter and $\mathbf{p}_\mrm{max}$ is the maximally allowed
three-momentum of the hadrons in their rest frame, left as a (in principle)
free parameter.
We don't have access to $b$ for each pair, but assume it is the same
distribution for all combinations of hadron types. A common restriction 
on the three-momentum is thus introduced for all pairs, which is implemented
as a cut on the invariant mass of the hadron pair,
\begin{equation}
m_\mathrm{inv} < \sqrt{m_1^2+|\mathbf{p}_\mrm{max}|^2}+
\sqrt{m_2^2+|\mathbf{p}_\mrm{max}|^2}~,
\label{eq:mCut}
\end{equation}
with $m_1$ and $m_2$ being the masses of the hadrons and $m_\mathrm{inv}$ the
physical invariant mass of the hadron pair. For all hadron
pairs that are not excluded by \eqRef{eq:mCut} we calculate the difference
in rapidity, $\Delta y=|y_1-y_2|$, and the rescattering probability.
For hadrons that are \textit{not} produced in the same string the latter is
\begin{equation}
P_\mathrm{ds}(\Delta y)=P^\mathrm{max}_{\,\mathrm{ds}}\left(1-\frac{\Delta y}
{\Delta y^\mathrm{max}}\right)~,
\label{eq:scatProbDS}
\end{equation}
where the maximum scattering probability $P^\mathrm{max}_{\,\mathrm{ds}}$ 
and the maximum rapidity difference $\Delta y^\mathrm{max}$ are left as free 
parameters. \EqRef{eq:scatProbDS} simply means a probability of 
$P^\mathrm{max}_{\,\mathrm{ds}}$ for zero rapidity difference of the hadron
pair, linearly decreasing to zero at a rapidity difference of 
$\Delta y^\mathrm{max}$.

As an alternative to \eqRef{eq:scatProbDS}, without the cut on the invariant
mass in \eqRef{eq:mCut}, the probability can be chosen to be
\begin{equation}
P_\mathrm{ds}(\Delta y,\Delta\varphi)=P^\mathrm{max}_{\,\mathrm{ds}}\left(1-
\frac{\sqrt{(\Delta y)^2+(\Delta\varphi)^2}}{R^\mathrm{max}}\right)~,
\label{eq:scatProbDSalt}
\end{equation}
with $\Delta\varphi$ being the difference in azimuth of the hadron pair and
$R^\mrm{max}$ the maximally allowed value of the radius $R=\sqrt{(\Delta y)^2
+(\Delta\varphi)^2}$.

For hadron pairs that are produced in the \textit{same} string we 
introduce the difference in hadron index (called rank e.g. in 
\cite{Field:1977fa}), $\Delta_{ij}=|i-j|$, to denote how 
close two hadrons are, i.e. two neighbours have $\Delta_{ij}=\Delta_{i~i+1}=1$,
next-to-neighbours $\Delta_{ij}=\Delta_{i~i+2}=2$ and so on. 
The scattering probability for same-string hadrons is
\begin{equation}
P_\mathrm{ss}(\Delta y)=P_\mathrm{ds}(\Delta y)\cdot\left\{
\begin{matrix*}[l]
P^\mathrm{max}_{\,\mathrm{ss}} & 
\text{if }\Delta_{ij}>\Delta_{ij}^\mathrm{max} \\[2mm]
\dfrac{P^\mathrm{max}_{\,\mathrm{ss}}(\Delta_{ij}-\Delta_{ij}^\mathrm{min})
+P^\mathrm{min}_{\,\mathrm{ss}}(\Delta_{ij}^\mathrm{max}-\Delta_{ij})}
{\Delta_{ij}^\mathrm{max}-\Delta_{ij}^\mathrm{min}} & 
\text{if }\Delta_{ij}^\mathrm{min}\le\Delta_{ij}\le\Delta_{ij}^\mathrm{max} 
\\[2mm]
0 & \text{if }\Delta_{ij}<\Delta_{ij}^\mathrm{min}~,
\end{matrix*}
\right.
\label{eq:scatProbSS}
\end{equation}
where $P_{\,\mathrm{ss}}^\mathrm{min/max}$ is the minimum/maximum
probability associated with the nearest/furthest neighbour, characterized
by $\Delta_{ij}^\mathrm{min/max}$, with a linear behaviour of the 
probability in between; zero probability for hadrons closer than 
$\Delta_{ij}^\mathrm{min}$ and maximum probability for those further
apart than $\Delta_{ij}^\mathrm{max}$. All four are left as 
free parameters. In the case where \eqRef{eq:scatProbDSalt} is applied,
$P_\mathrm{ds}(\Delta y)$ in \eqRef{eq:scatProbSS} has to be replaced by 
$P_\mathrm{ds}(\Delta y,\Delta\varphi)$.

\subsubsection{Multi-string toy model}

\begin{figure}[t!]
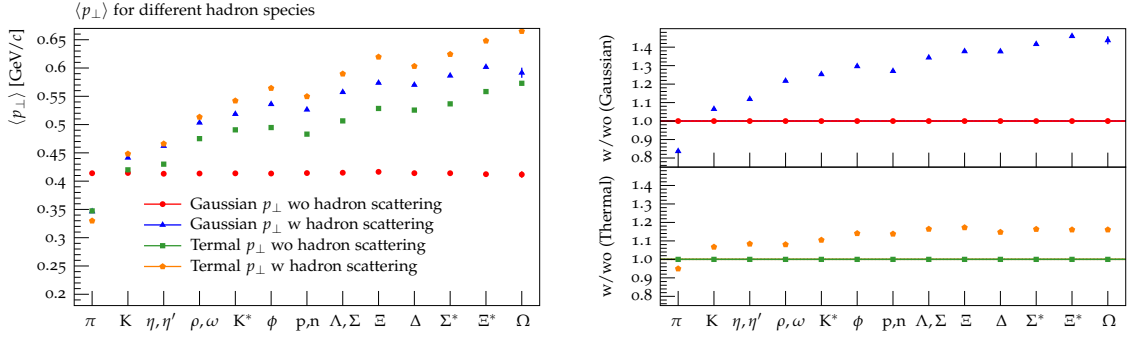

\centering
\begin{subfigure}[b]{0.49\textwidth}
  \rivetFigure{hadronScatterToyModel/avPTtoyCollFlow.pdf}
\end{subfigure}
\hfill
\begin{subfigure}[b]{0.49\textwidth}
  \rivetFigureRatio{hadronScatterToyModel/avPTtoyCollFlow-ref1.pdf} 
  \rivetFigure{hadronScatterToyModel/avPTtoyCollFlow-ref2.pdf}
\end{subfigure}
\\
\caption{The $\langle p_\perp \rangle$ for different hadron species,
with (blue triangles) and without (red dots) hadron scattering for the 
Gaussian $p_\perp$ and with (orange pentagons) and without (green squares) 
hadron scattering for the thermal $p_\perp$ in the toy model. The right
plots show the ratio of the Gaussian model with to without hadron 
scattering in the upper panel and the same ratio
for the thermodynamical model in the lower panel.
\label{fig:avPTtoy}}
\end{figure}

\begin{figure}[p!]
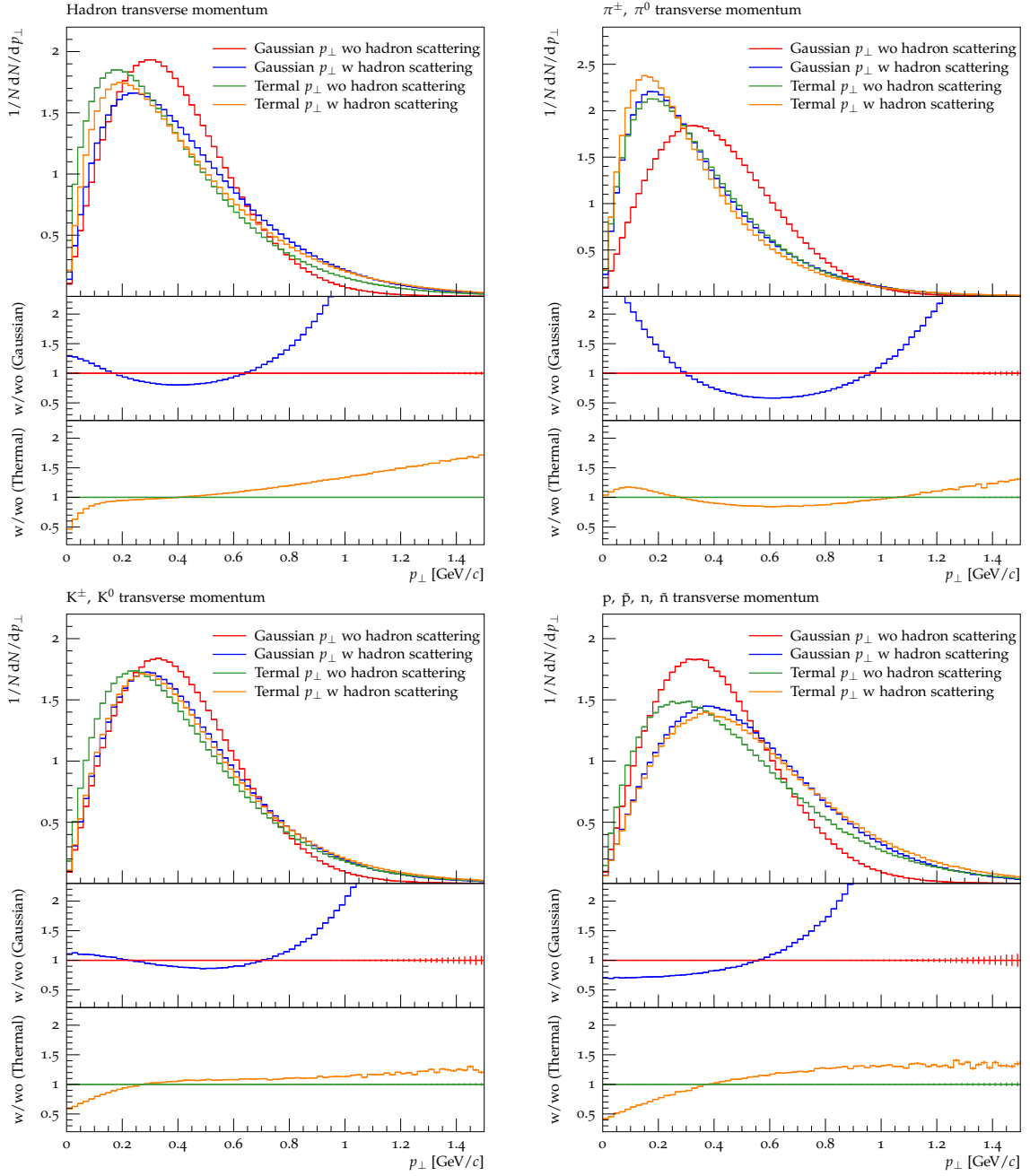

\centering
\rivetFigure{hadronScatterToyModel/pThadToyCollFlow.pdf} \hfill
\rivetFigureRatio{hadronScatterToyModel/pTpiToyCollFlow.pdf} \\
\rivetFigure{hadronScatterToyModel/pThadToyCollFlow-ref1.pdf} \hfill
\rivetFigureRatio{hadronScatterToyModel/pTpiToyCollFlow-ref1.pdf} \\
\rivetFigure{hadronScatterToyModel/pThadToyCollFlow-ref2.pdf} \hfill
\rivetFigureRatio{hadronScatterToyModel/pTpiToyCollFlow-ref2.pdf} \\
\rivetFigure{hadronScatterToyModel/pTkToyCollFlow.pdf} \hfill
\rivetFigureRatio{hadronScatterToyModel/pTpnToyCollFlow.pdf} \\
\rivetFigure{hadronScatterToyModel/pTkToyCollFlow-ref1.pdf} \hfill
\rivetFigureRatio{hadronScatterToyModel/pTpnToyCollFlow-ref1.pdf} \\
\rivetFigure{hadronScatterToyModel/pTkToyCollFlow-ref2.pdf} \hfill
\rivetFigure{hadronScatterToyModel/pTpnToyCollFlow-ref2.pdf} \\
\caption{The $p_\perp$ spectrum for all hadrons (\textit{top left}), pions
(\textit{top right}), kaons (\textit{bottom left}) and protons and neutrons 
(\textit{bottom right}), with (blue) and without (red) hadron scattering 
for the Gaussian $p_\perp$ and with (orange) and without (green) hadron 
scattering for the thermal $p_\perp$ in the toy model.
\label{fig:hadPTtoy}}
\end{figure}

In order to test the hadron scattering a simple toy model is applied:
five strings, each with energy $\mZ$, are constructed along the $z$
axis with different quark flavours for the endpoint quarks, and the 
primary produced hadrons are studied.
As in \secRef{sec:oldModelVar} the hadrons containing the endpoint 
quarks are excluded.
The following plots are obtained by making use of \eqsRef{eq:mCut} and
\eqref{eq:scatProbDS}; using \eqRef{eq:scatProbDSalt} instead leads
to similar results, and thus the same conclusions.

\FigRef{fig:avPTtoy} shows the $\langle p_\perp \rangle$ for
different hadron species. For the Gaussian hadronic $p_\perp$ distribution 
without hadron scattering all hadrons receive the same $p_\perp$ spectrum. 
Including hadron scattering, the $\langle p_\perp \rangle$ decreases for pions, 
the lightest hadrons, by about $20\%$ and increases for heavy hadrons by up 
more than $40\%$. The same effect is present for the thermodynamical model, 
although changes only reach around $10\%$, as the $\langle p_\perp \rangle$ is 
higher for heavier hadrons already without hadron rescattering.

\FigRef{fig:hadPTtoy} shows the normalized $p_\perp$ spectra for 
all hadrons and, to exemplify the difference between light and heavy hadrons, 
the spectrum for pions, kaons, and protons/neutrons. Comparing the inclusive 
$p_\perp$ spectrum for the Gaussian $p_\perp$, we notice that the distribution 
gets broader with hadron scattering, i.e. we get more pions with small 
$p_\perp$ and more heavy hadrons with higher $p_\perp$.

As the thermodynamical model without hadron rescattering comes with a different 
starting point, compared to the Gaussian model, the effect of the rescattering
is not as large, and shifts the $p_\perp$ spectrum towards larger values. 
The exception is for pions only, where there is a slight broadening also 
towards smaller $p_\perp$ values.

\subsection{Results}

\subsubsection{The Effect of Decays}
\label{sec:effectDecays}

\begin{figure}[p!]
\centering
\rivetFigure{effectOfDecays/pTpi-LEP.pdf} \hfill
\rivetFigureRatio{effectOfDecays/pTp-LEP.pdf} \\
\rivetFigure{effectOfDecays/pTpi-LEP-ref1.pdf} \hfill
\rivetFigureRatio{effectOfDecays/pTp-LEP-ref1.pdf} \\
\rivetFigure{effectOfDecays/pTpi-LEP-ref2.pdf} \hfill
\rivetFigureRatio{effectOfDecays/pTp-LEP-ref2.pdf} \\
\rivetFigure{effectOfDecays/pTpi-LEP-ref3.pdf} \hfill
\rivetFigureRatio{effectOfDecays/pTp-LEP-ref3.pdf} \\
\rivetFigure{effectOfDecays/pTpi-LEP-ref4.pdf} \hfill
\rivetFigure{effectOfDecays/pTp-LEP-ref4.pdf} \\
\rivetFigure{effectOfDecays/pTpi-LHC.pdf} \hfill
\rivetFigureRatio{effectOfDecays/pTp-LHC.pdf} \\
\rivetFigure{effectOfDecays/pTpi-LHC-ref1.pdf} \hfill
\rivetFigureRatio{effectOfDecays/pTp-LHC-ref1.pdf} \\
\rivetFigure{effectOfDecays/pTpi-LHC-ref2.pdf} \hfill
\rivetFigureRatio{effectOfDecays/pTp-LHC-ref2.pdf} \\
\rivetFigure{effectOfDecays/pTpi-LHC-ref3.pdf} \hfill
\rivetFigureRatio{effectOfDecays/pTp-LHC-ref3.pdf} \\
\rivetFigure{effectOfDecays/pTpi-LHC-ref4.pdf} \hfill
\rivetFigure{effectOfDecays/pTp-LHC-ref4.pdf} \\
\caption{The $p_\perp$ spectrum for charged pions (\textit{left})
and protons (\textit{right}), with (blue) and without (red) hadronic 
decays for the Gaussian and with (orange) and without (green) hadronic 
decays for the thermal model in $\e^+\e^-\to$~jets (\textit{top}) 
and inelastic $\p\p$ collisions (\textit{bottom}).
\label{fig:hadPTdecays}}
\end{figure}

Hadron decays, such as $\rho\to\pi\pi$ or $\eta\to\pi^+\pi^-\pi^0$,
influence the $\pT$ spectra of the final state hadrons. As the decays 
are mostly dictated by kinematics, they constitute a limiting factor 
on the possibilities of modifying for instance the pion $\pT$ spectrum
during the fragmentation process. Even though the primary hadrons
follow a Gaussian or exponential $\pT$ distribution, the spectra 
obtained after decays do not, and become more similar. In addition,
in realistic events the effects of perturbative jet production leads
to a $\pT$ broadening and the emergence of a powerlike high-$\pT$
tail.

To investigate this smearing of the $\pT$ spectra, we consider realistic 
$\e^+\e^-\to$~jets events and inelastic $\p\p$ collisions, where the 
previously discussed effects of string density and hadron rescattering are 
not taken into account. The normalized transverse momentum distributions
of $\pi^\pm$ and $\p,\bar\p$ are shown in \figRef{fig:hadPTdecays}
for the Gaussian and the thermodynamical model. Comparing the $\pT$
spectra to the previous plots (note the different range of $x$ and 
$y$ axis) reveals how much the $\pT$ distributions have moved to
higher $\pT$ values, as mentioned above. Four different ratio 
plots are included for each histogram to investigate different effects:
the ratio of distributions after hadronic decays with respect to before,
for both models, and the ratio of the thermodynamical with respect to 
the Gaussian model, with and without decays.

Decays shift the $\pT$ spectra towards smaller values, where 
the Gaussian model shows a larger change, compared to the thermodynamical
model. For LEP the difference between the predictions with and without 
decays is limited to around $50\%$ at most, while for LHC the changes 
are rather large, especially for small and large $\pT$ values.
\FigRef{fig:hadPTdecays} also nicely shows that the differences between the 
Gaussian and thermodynamical model become more than a factor of two smaller
when hadronic decays are included, where, as before, the differences are
more pronounced for LHC events. Unfortunately, this will limit the impact 
of the modifications previously discussed in this section.

\begin{figure}[t!]
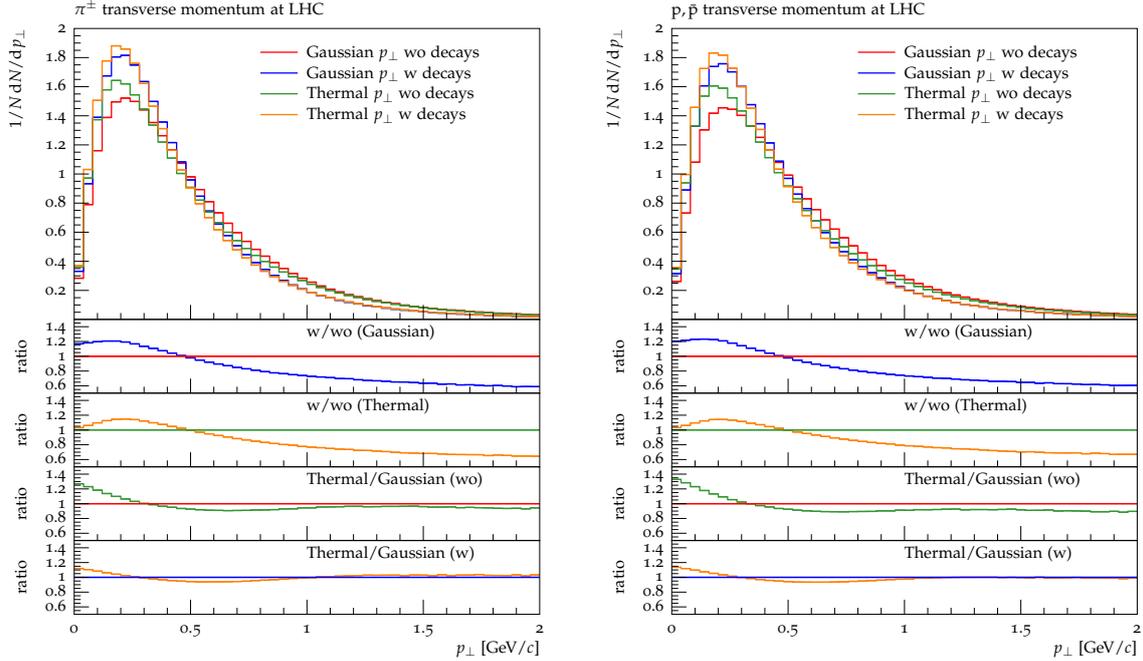

\centering
\rivetFigure{effectOfDecays/pTpi-LHC-HS.pdf} \hfill
\rivetFigureRatio{effectOfDecays/pTp-LHC-HS.pdf} \\
\rivetFigure{effectOfDecays/pTpi-LHC-HS-ref1.pdf} \hfill
\rivetFigureRatio{effectOfDecays/pTp-LHC-HS-ref1.pdf} \\
\rivetFigure{effectOfDecays/pTpi-LHC-HS-ref2.pdf} \hfill
\rivetFigureRatio{effectOfDecays/pTp-LHC-HS-ref2.pdf} \\
\rivetFigure{effectOfDecays/pTpi-LHC-HS-ref3.pdf} \hfill
\rivetFigureRatio{effectOfDecays/pTp-LHC-HS-ref3.pdf} \\
\rivetFigure{effectOfDecays/pTpi-LHC-HS-ref4.pdf} \hfill
\rivetFigureRatio{effectOfDecays/pTp-LHC-HS-ref4.pdf} \\
\caption{The $p_\perp$ spectrum for charged pions (\textit{left})
and protons (\textit{right}), with (blue) and without (red) hadronic 
decays for the Gaussian and with (orange) and without (green) hadronic 
decays for the thermal model in inelastic $\p\p$ collisions
(\textit{bottom}) with hadron rescattering.
\label{fig:hadPTdecaysHS}}
\end{figure}

Another question is how much of the hadron rescattering effect 
on the primary hadrons survives the decays. Note that some of the primary 
hadrons, such as the $\rho$ meson, are so short-lived that some of their 
decay products could rescatter which would influence the $\pT$ spectra further. 
A realistic interleaving of rescattering and decays would require a
detailed space--time picture, however, which is for the future.

In \figRef{fig:hadPTdecaysHS} the pion and proton $\pT$ spectra are 
shown again for LHC events with hadron rescattering. The same ratio 
plots as before are included.
The ratio of predictions with to without hadron decays shows a similar 
behaviour as the plots before, where no rescattering was included.
The difference between the two models without hadronic decays becomes smaller
when hadron rescattering is included. This it not surprising since the effect 
of the rescattering, that of shuffling some $\pT$ from 
lighter to heavier hadrons, is smaller in the thermodynamical model where 
more massive hadrons obtain more $\pT$ already form the beginning. Including 
decays brings the predictions of the two models even closer together.

\subsubsection{Adding more Hadrons}
\label{sec:moreHadrons}

We now briefly investigate the effect of including additional hadrons
in the flavour picking process in the thermodynamical model.
$\ee\to$~jets events at $\sqrt{s}=m_\Z$ are analyzed with the effects 
of string density and hadron rescattering not being used. As discussed 
in \secRef{sec:thermModel}, by default only hadrons with $\u/\d/\s$ 
quarks and no radial or orbital excitation are included.

\begin{figure}[t!]
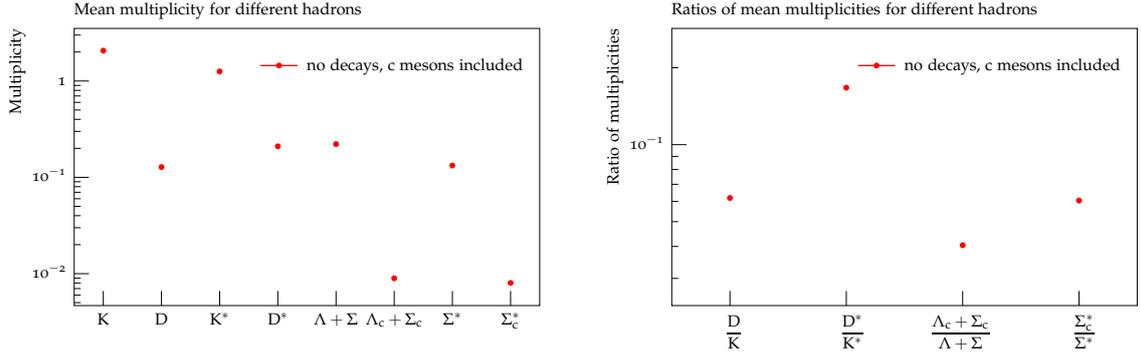

\centering
\rivetFigure{effectOfMoreHadrons/meanCnonCMulti.pdf} \hfill
\rivetFigure{effectOfMoreHadrons/relMeanCnonCMulti.pdf} \\
\caption{Mean multiplicities and their ratios for different groups of
hadrons. Predictions of the thermodynamical model with $\c$ hadrons
included in the fragmentation process and no hadronic decays.
\label{fig:cHadrons}}
\end{figure}

Firstly, consider including hadrons with charm quarks. To obtain a rough 
estimate of the suppression of $\c$ production in string breaks, compared 
to that of $\s$ quarks, the rates of $\D$ and $\K$ mesons
and their ratios are analyzed; similar for vector mesons and baryons. 
The results in \figRef{fig:cHadrons} show that the $\c$ hadrons are 
suppressed by more than an order of magnitude compared to $\s$ hadrons, 
although a bit less when only vector mesons are considered. In absolute
numbers the amount of extra charm production is non-negligible, and probably 
inconsistent with both LEP and LHC observed rates. Recall that an
additional suppression factor for $\s$ quarks was introduced for the hadron 
rates in \secRef{sec:thermModel}; we would therefore expect that a similar, 
even stronger factor is needed when including $\c$ quarks in the 
thermodynamical model. Given this, neither charm nor bottom production is 
included in the nonperturbative hadronization in the rest of our studies.

\begin{figure}[t!]
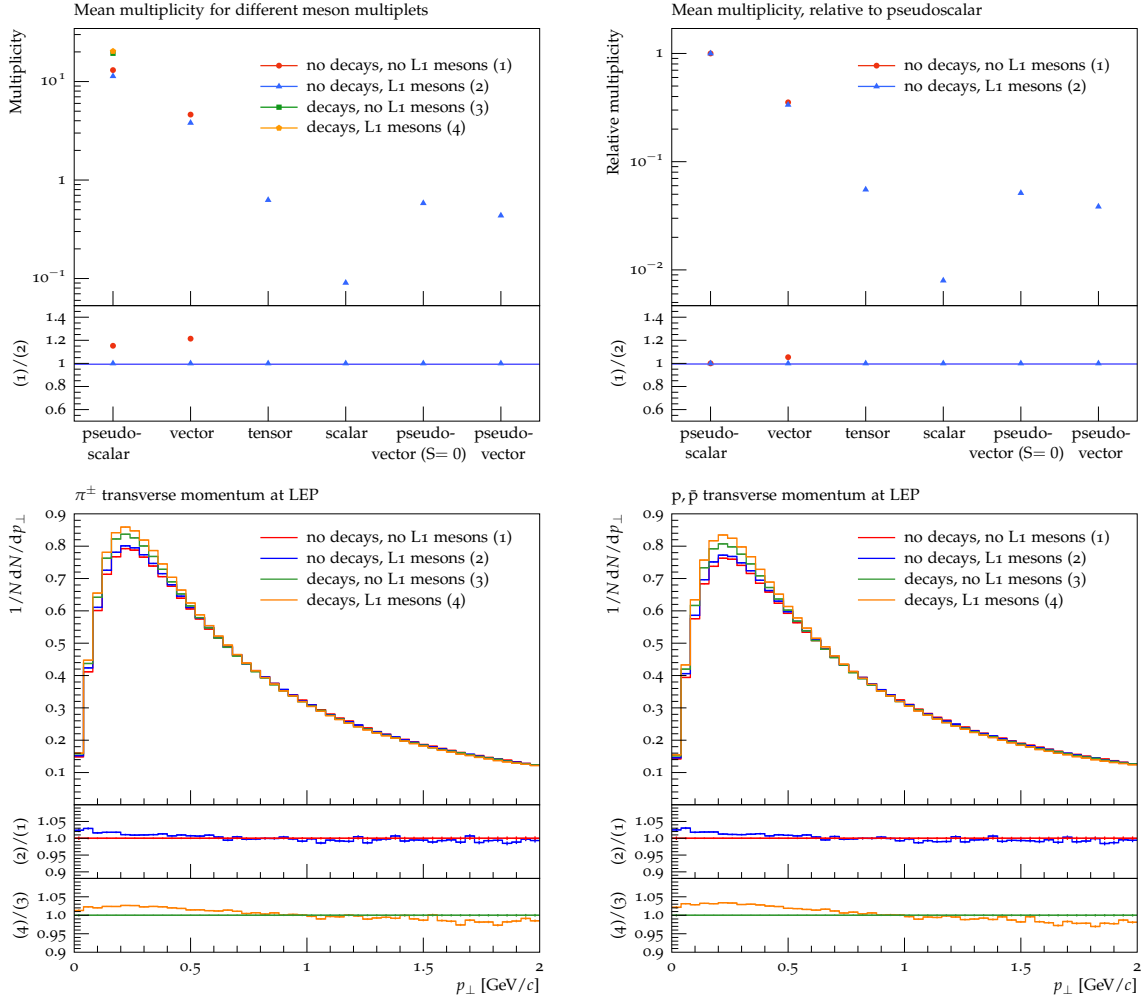

\centering
\rivetFigure{effectOfMoreHadrons/meanMesonMulti.pdf} \hfill
\rivetFigure{effectOfMoreHadrons/relMeanMesonMulti.pdf} \\
\rivetFigure{effectOfMoreHadrons/pTpi.pdf} \hfill
\rivetFigureRatio{effectOfMoreHadrons/pTp.pdf} \\
\rivetFigure{effectOfMoreHadrons/pTpi-ref1.pdf} \hfill
\rivetFigureRatio{effectOfMoreHadrons/pTp-ref1.pdf} \\
\rivetFigure{effectOfMoreHadrons/pTpi-ref2.pdf} \hfill
\rivetFigureRatio{effectOfMoreHadrons/pTp-ref2.pdf} \\
\caption{Mean multiplicities and ratios for the different
meson multiplets (\textit{top}).
The $p_\perp$ spectrum for charged pions (\textit{bottom left})
and protons (\textit{bottom right}).
Predictions of the thermodynamical are shown, where decays are 
either switched on or off and $L=1$ mesons are included or not.
\label{fig:excHadrons}}
\end{figure}

Secondly, consider excited mesons by including all of the following meson 
multiplets, where by default only the first two are present,
\begin{Itemize}
\item pseudoscalar multiplet with $L=0,~S=0,~J=0~,$
\item vector multiplet with $L=0,~S=1,~J=1~,$
\item pseudovector multiplet with $L=1,~S=0,~J=1~,$
\item scalar multiplet with $L=1,~S=1,~J=0~,$
\item pseudovector multiplet with $L=1,~S=1,~J=1~,$
\item tensor multiplet with $L=1,~S=1,~J=2~,$
\end{Itemize}
with $J$ denoting the sum of the spin $S$ and orbital angular momentum $L$
in the nonrelativistic approximation.
In \figRef{fig:excHadrons} the mean multiplicity of the different multiplets
is shown, together with the $\pT$ spectra of pions and protons. Note
that including excited mesons leads to an increase of the total meson 
multiplicity after decays. All $L=1$ multiplets are suppressed by more than 
an order of magnitude with respect to the pseudoscalar multiplet, with the 
scalar mesons being suppressed the most due the combination of them being 
the heaviest of the considered hadrons and their smaller spin-state weight 
$2J+1$. The normalized $\pT$ spectra exhibit slight shifts towards smaller 
values, as the now included heavier mesons decay to more lighter hadrons.
The excited mesons combined constitute a fraction of roughly $10\%$ of the 
total meson multiplicity. Given that in addition those mesons and their 
decay channels are not very well understood, we consider it reasonable 
to not include those in further studies. In default \textsc{Pythia} the 
suppression of light vector mesons with respect to pseudoscalar mesons 
is $\sim0.5$. The thermodynamical naturally comes with a fairly similar 
value of $\sim0.35$. 

\section{Comparisons with Data}
\label{sec:results}

We now proceed to compare the models with data. Note that, in a first step,
there is no ambition to obtain a better overall description than the one 
achieved in several of the standard tunes that come with \textsc{Pythia}. 
It is rather to explore how the modelling of the new mechanisms impacts 
selected distributions, notably the ones discussed in \secRef{sec:keydata}. 
That is, whether the mechanisms have the potential to improve the
agreement with data in some crucial respects. Only in a second step 
is there some attempt to combine the various mechanisms, but still without
the ambition of a full-fledged tune. 
In \secRef{sec:individualResults} we present a comparison of the different
effects we have discussed so far, while \secRef{sec:finalResults}
gives an overview of the results obtained by combining the effects into
a more complete picture.
Note that the new mechanisms will be available in the next public 
\textsc{Pythia} release.

\subsection{Impact of the Different Effects}
\label{sec:individualResults}

\begin{figure}[p!]
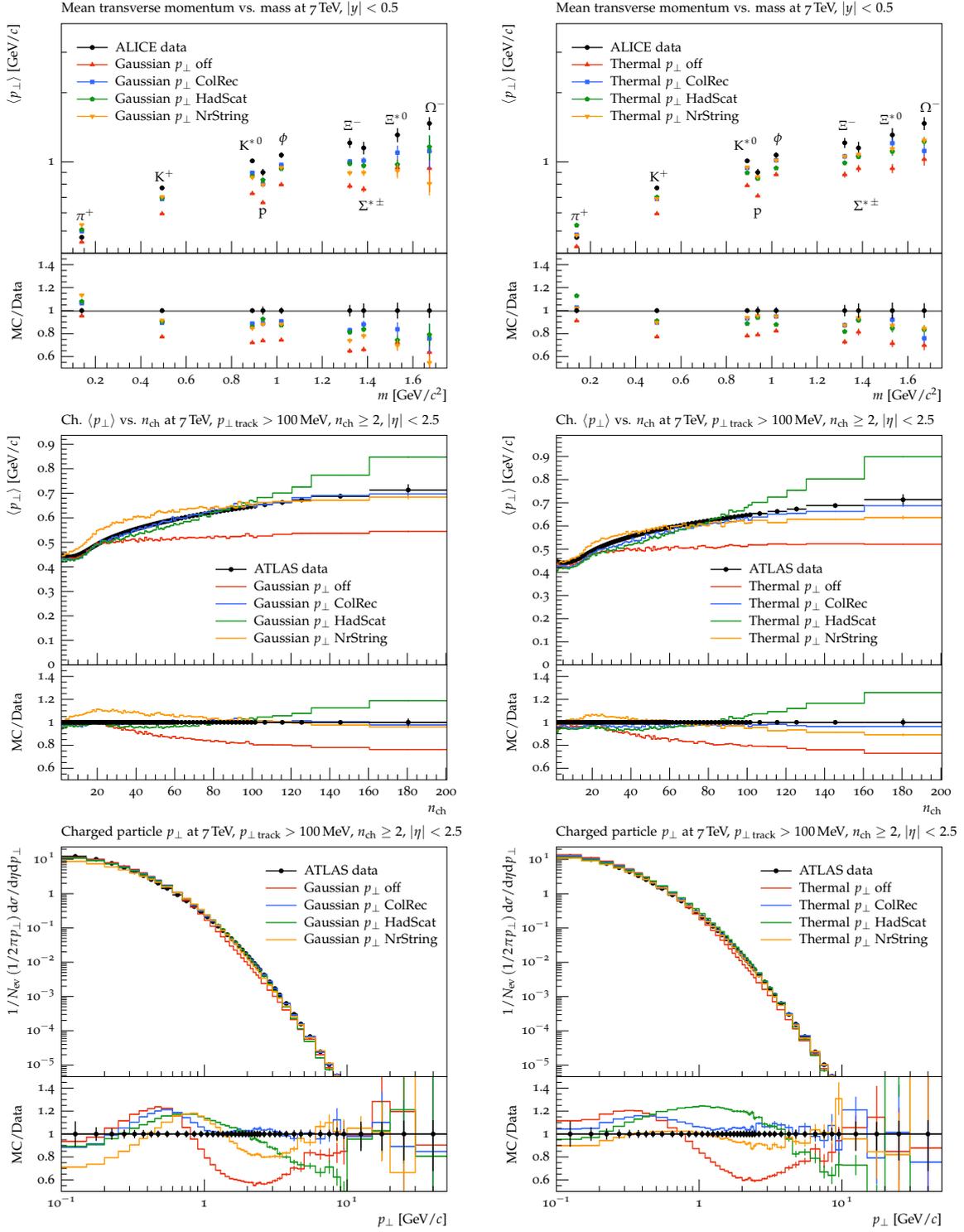

\centering
\rivetFigure{data/ALICE_2014_I1300380/d03-x01-y01-pTm-checkOld.pdf} \hfill
\rivetFigure{data/ALICE_2014_I1300380/d03-x01-y01-pTm-checkNew.pdf} \\
\rivetFigure{data/ATLAS_2010_S8918562/d25-x01-y01-pTnCh-checkOld.pdf} \hfill
\rivetFigure{data/ATLAS_2010_S8918562/d25-x01-y01-pTnCh-checkNew.pdf} \\
\rivetFigure{data/ATLAS_2010_S8918562/d12-x01-y01-pThad-checkOld.pdf} \hfill
\rivetFigure{data/ATLAS_2010_S8918562/d12-x01-y01-pThad-checkNew.pdf} \\
\caption{Comparisons to ALICE~\cite{Abelev:2014qqa} and 
ATLAS~\cite{Aad:2010ac} data: $\meanPT$ as a function of the hadron 
mass (\textit{top}), charged multiplicity (\textit{middle}), and the 
charged particle $\pT$ (\textit{bottom}). Predictions with the Gaussian 
(thermodynamical) model are shown in the \textit{left} (\textit{right}) 
plots. ColRec\,/\,HadScat\,/\,NrString means that only colour 
reconnection\,/\,hadron rescattering\,/\,$n_\mrm{string}^\mrm{eff}$-dependence
is switched on, otherwise everything is switched off.
\label{fig:resEff}}
\end{figure}

Based on a limited set of LHC observables, this section presents the impact
of the new mechanisms outlined in \secRef{sec:newModels} on the description of 
data. The observables have been chosen to illustrate the effects of the 
change of the Gaussian width or temperature, respectively, depending on the 
close-packing of strings as in \eqsRef{eq:nStringEffKappa} and 
\eqref{eq:nStringEffTemp}, of hadron rescattering, and of colour reconnection. 
The latter has been included as it serves a similar purpose and shows a 
somewhat comparable behaviour. The baseline prediction, which serves as the 
main comparison for both models, is obtained by switching off all of the 
aforementioned effects. For a clear picture of the influence of the 
individual mechanisms, only one of them is switched on at a time. Note that 
the prediction of the Gaussian model with colour reconnection, labelled 
``Gaussian $\pT$ ColRec'' in the plots, corresponds to default 
\textsc{Pythia}~8. Recall that the results presented in this subsection are 
not obtained with parameter settings that optimize the data description but 
rather illustrate their general effect. The average transverse momentum 
$\meanPT$ as a function of the hadron mass and the charged multiplicity 
are shown in \figRef{fig:resEff}, together with the charged particle $\pT$ 
spectrum.

For both models, the description of $\meanPT$ as a function of 
mass improves for each of the different mechanisms, compared to the baseline 
prediction, as heavier hadrons obtain larger $\meanPT$ values. 
The thermodynamical model provides a somewhat better description of 
this observable, compared to the Gaussian model, which comes naturally due 
to the exponential hadronic transverse-mass suppression.

The baseline prediction for $\langle\pT\rangle(n_\mrm{ch})$ plateaus at
small multiplicities, therefore underestimating $\meanPT$ for 
values $n_\mrm{ch}\gtrsim25$. All of the effects investigated in this study
have a somewhat similar effect, in the sense that they are able to push
up the prediction, compared to the baseline settings. While including the 
$n_\mrm{string}^\mrm{eff}$-dependence significantly improves the description,
it is still slightly worse than the prediction with colour reconnection. 
The hadron rescattering provides a fairly good description of 
$\meanPT$ for small $n_\mrm{ch}$ values, but clearly overshoots 
the distribution at high multiplicities.

Similar to the previous observable, the $n_\mrm{string}^\mrm{eff}$-dependence
and colour reconnection improve the description of the inclusive $\pT$ 
spectrum. The Gaussian model without additional effects switched on produces 
a bump at $\pT\sim 0.5~\mrm{GeV}/c$ and a broad dip at $\pT\sim 2.5~
\mrm{GeV}/c$. While colour reconnection removes the dip almost completely, the 
bump is still clearly visible. The $n_\mrm{string}^\mrm{eff}$-dependence 
somewhat reduces both bump and dip, but at the cost of introducing another dip 
towards very small $\pT$ values. The baseline prediction of the 
thermodynamical model, compared with the Gaussian one, has the same dip at 
$\pT\sim 2.5~\mrm{GeV}/c$, while the bump is much less visible.
Both colour reconnection and the $n_\mrm{string}^\mrm{eff}$-dependence 
reduce the dip quite substantially and provide a very good description of the 
data. The hadron rescattering, while somewhat improving the description in the 
low-$\pT$ region, overestimates mid-$\pT$ values by around $20\%$
before undershooting the distribution.

\subsection{Results}
\label{sec:finalResults}

Using the information of the last subsection we adjust the parameters 
associated with the new mechanisms to obtain a good data description, with 
the main focus lying on $\pT$ spectra of pions, kaons, and protons. We begin 
with LHC data, as this is the motivation for the thermodynamical model, the 
$n_\mrm{string}^\mrm{eff}$-dependence, and the hadron rescattering, and 
continue with a cross-check of some LEP and SLC observables.

\subsubsection{LHC}

\begin{figure}[p!]
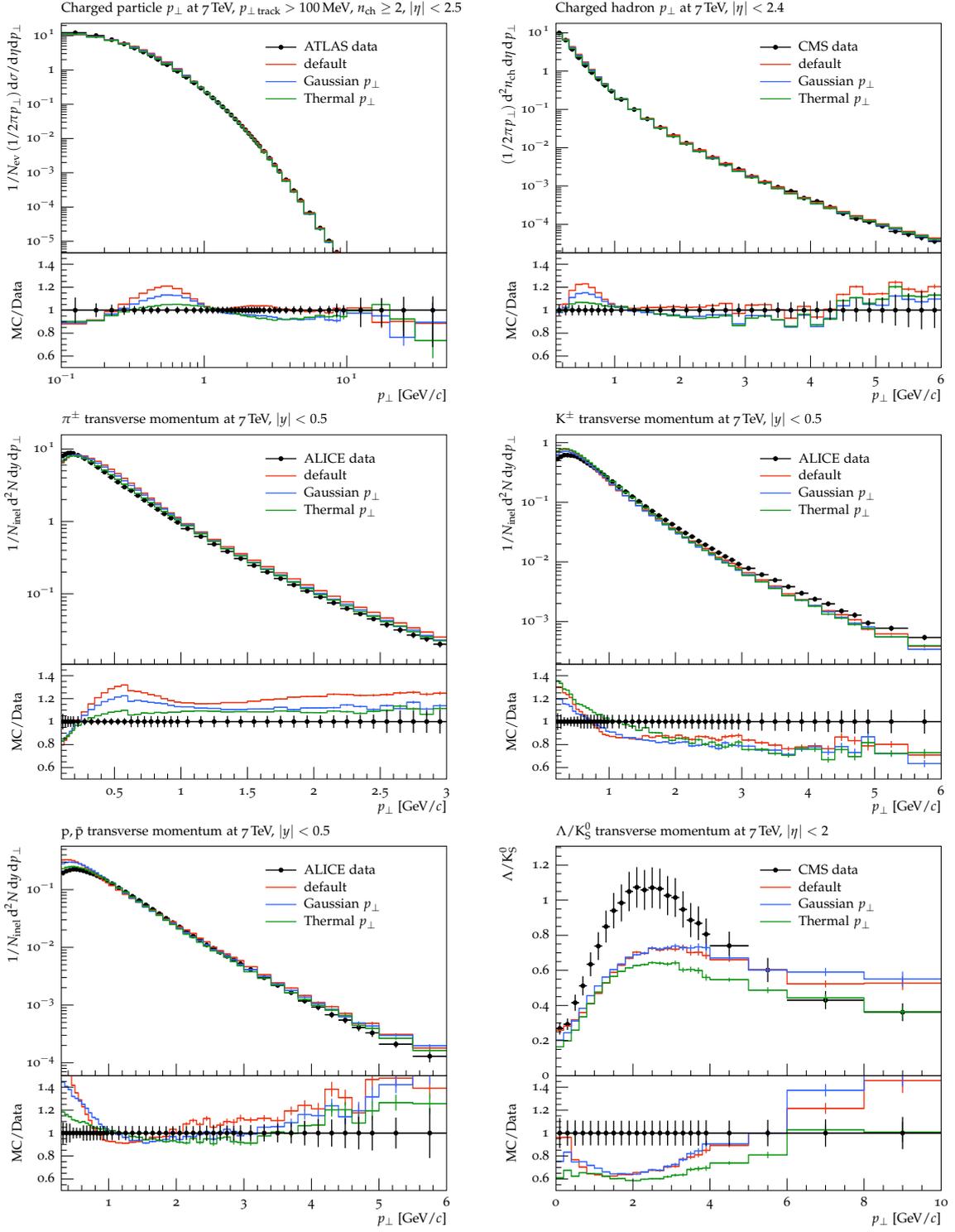

\centering
\rivetFigure{data/ATLAS_2010_S8918562/d12-x01-y01-pThad.pdf} \hfill
\rivetFigure{data/CMS_2010_S8656010/d04-x01-y01-pThad.pdf} \\
\rivetFigure{data/ALICE_2015_I1357424/d01-x01-y01-pTpi.pdf} \hfill
\rivetFigure{data/ALICE_2015_I1357424/d01-x01-y02-pTk.pdf} \\
\rivetFigure{data/ALICE_2015_I1357424/d01-x01-y03-pTp.pdf} \hfill
\rivetFigure{data/CMS_2011_S8978280/d07-x01-y02-pTlambdaKs0.pdf} \\
\caption{Inclusive (\textit{top}), $\pi^\pm$, $\K^\pm$, and $\p,\pbar$ 
(\textit{middle} and \textit{bottom left}) $\pT$ spectra and the $\Lambda$ to
$\K_\mathrm{S}^0$ ratio (\textit{bottom right}). Predictions of default 
\textsc{Pythia}, the Gaussian and thermodynamical model with modifications, 
compared to ATLAS~\cite{Aad:2010ac}, CMS~\cite{Khachatryan:2010us,
Khachatryan:2011tm} and ALICE data~\cite{Adam:2015qaa}.
\label{fig:LHCpTspectra}}
\end{figure}

\begin{figure}[p!]
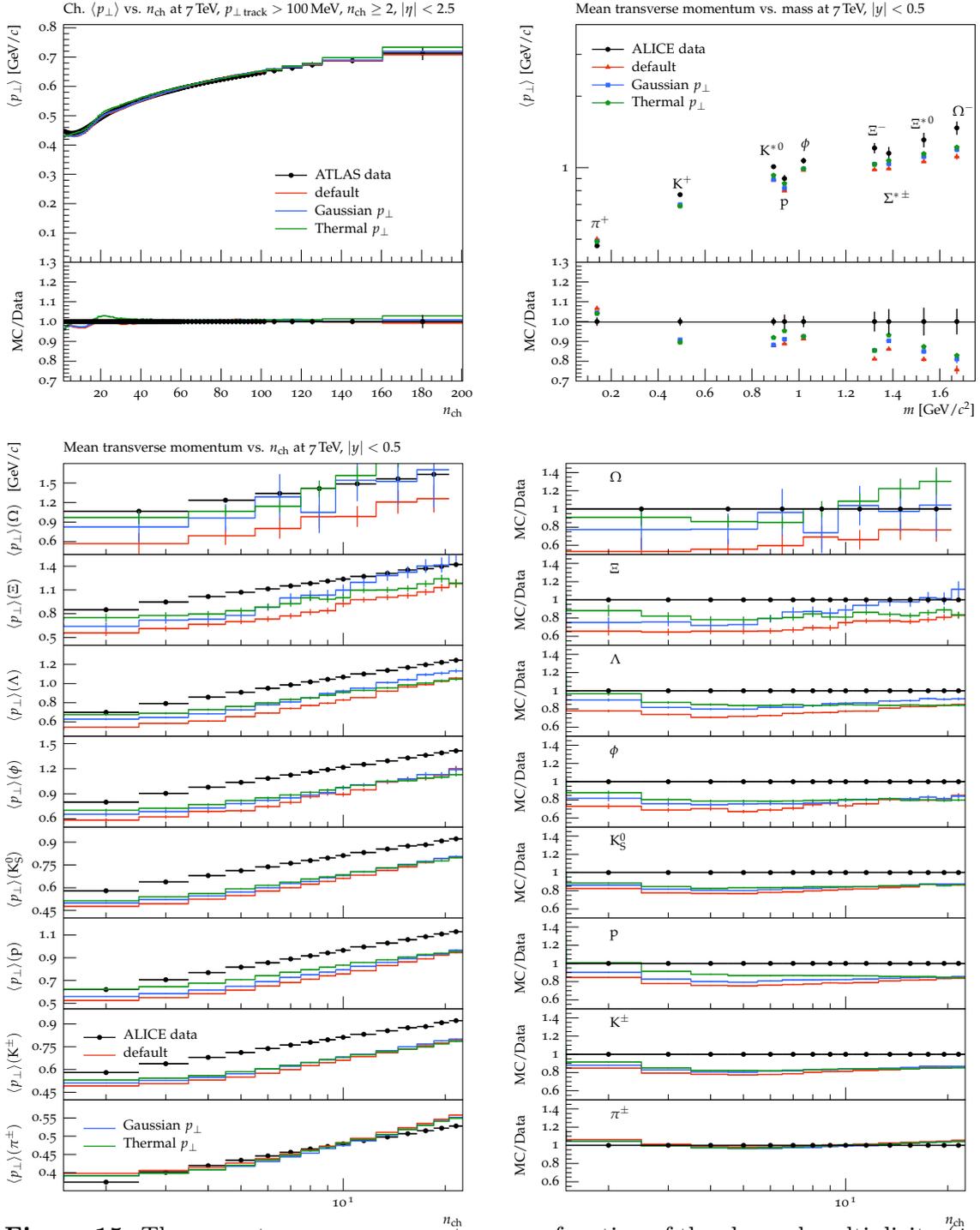

\centering
\rivetFigure{data/ATLAS_2010_S8918562/d25-x01-y01-pTnCh.pdf} \hfill
\rivetFigure{data/ALICE_2014_I1300380/d03-x01-y01-pTm.pdf} \\[2mm]
\begin{subfigure}[b]{0.49\textwidth}
  \rivetFigureRatio{data/ALICE_2016/d02-x01-y08-pTnCh-Omega.pdf}
  \rivetFigureRatio{data/ALICE_2016/d02-x01-y07-pTnCh-Xi.pdf}
  \rivetFigureRatio{data/ALICE_2016/d02-x01-y06-pTnCh-Lambda.pdf}
  \rivetFigureRatio{data/ALICE_2016/d02-x01-y05-pTnCh-phi.pdf}
  \rivetFigureRatio{data/ALICE_2016/d02-x01-y04-pTnCh-ks.pdf}
  \rivetFigureRatio{data/ALICE_2016/d02-x01-y03-pTnCh-p.pdf}
  \rivetFigureRatio{data/ALICE_2016/d02-x01-y02-pTnCh-k.pdf}
  \rivetFigure{data/ALICE_2016/d02-x01-y01-pTnCh-pi.pdf}
\end{subfigure}
\hfill
\begin{subfigure}[b]{0.49\textwidth}
  \rivetFigureRatio{data/ALICE_2016/d02-x01-y08-pTnCh-Omega-ref.pdf}
  \rivetFigureRatio{data/ALICE_2016/d02-x01-y07-pTnCh-Xi-ref.pdf}
  \rivetFigureRatio{data/ALICE_2016/d02-x01-y06-pTnCh-Lambda-ref.pdf}
  \rivetFigureRatio{data/ALICE_2016/d02-x01-y05-pTnCh-phi-ref.pdf}
  \rivetFigureRatio{data/ALICE_2016/d02-x01-y04-pTnCh-ks-ref.pdf}
  \rivetFigureRatio{data/ALICE_2016/d02-x01-y03-pTnCh-p-ref.pdf}
  \rivetFigureRatio{data/ALICE_2016/d02-x01-y02-pTnCh-k-ref.pdf}
  \rivetFigure{data/ALICE_2016/d02-x01-y01-pTnCh-pi-ref.pdf}
\end{subfigure}
\\[-5mm]
\caption{The mean transverse momentum as a function of the charged 
multiplicity (\textit{top left}) and the hadron mass (\textit{top right}) 
and (\textit{bottom}). Predictions of default \textsc{Pythia}, the Gaussian 
and thermodynamical model with modifications, compared to 
ALICE~\cite{Bianchi:2016szl,Abelev:2014qqa} and ATLAS~\cite{Aad:2010ac} 
data. The data in the bottom plots is taken to be an estimate of the 
logarithmic fits in \cite{Bianchi:2016szl} and therefore no error bars
are included.
\label{fig:LHCpTvsMorNch}}
\end{figure}

\begin{figure}[tbp]
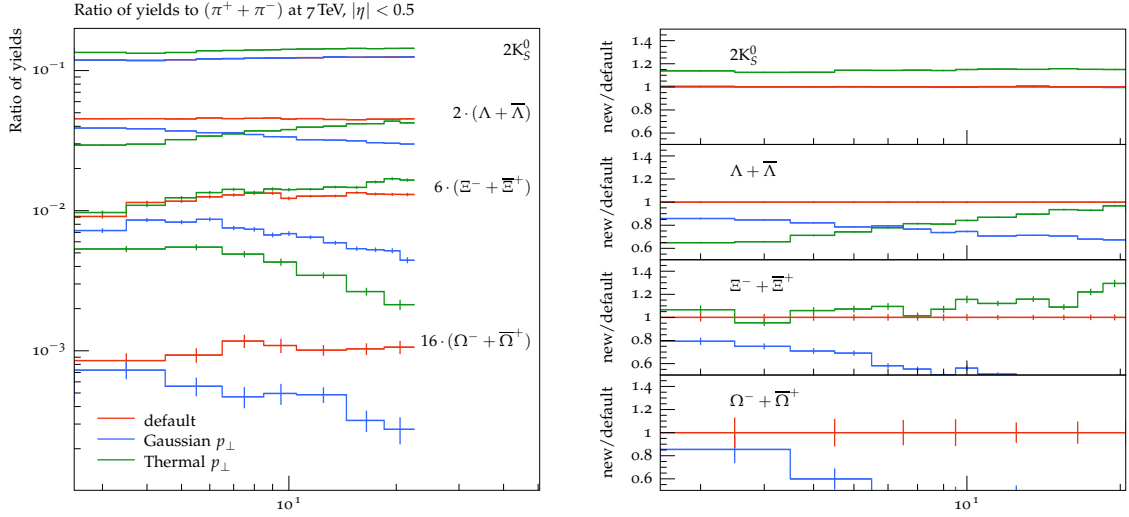

\centering
\begin{subfigure}[b]{0.49\textwidth}
  \rivetFigure{data/ALICE_2016/d01-x01-yieldRatios.pdf}
\end{subfigure}
\hfill
\begin{subfigure}[b]{0.49\textwidth}
  \rivetFigureRatio{data/ALICE_2016/d01-x01-yieldRatios-ref1.pdf}
  \rivetFigureRatio{data/ALICE_2016/d01-x01-yieldRatios-ref2.pdf}
  \rivetFigureRatio{data/ALICE_2016/d01-x01-yieldRatios-ref3.pdf}
  \rivetFigure{data/ALICE_2016/d01-x01-yieldRatios-ref4.pdf}
\end{subfigure}
\\[-5mm]
\caption{Ratio of yields with respect to $(\pi^++\pi^-)$ as a function of
the charged multiplicity. Predictions of default \textsc{Pythia}, the Gaussian 
and thermodynamical model with modifications. The ALICE measurement can be 
found in \cite{Adam:2016emw}.
\label{fig:LHCyieldRatios}}
\end{figure}

The new parameters are adjusted such that an improvement of the $\pT$
spectra of $\pi^\pm$, $\K^\pm$ and $\p,\bar\p$, measured with 
ALICE~\cite{Adam:2015qaa}, is achieved, while still giving a reasonable 
description of the charged particle $\pT$ distribution and 
$\meanPT$ as a function of the multiplicity, both measured
with ATLAS~\cite{Aad:2010ac}. The corresponding settings and values can be 
found in \appRef{app:settings}. The LHC data set presented here includes 
the aforementioned $\pT$ spectra and the $\Lambda$ to $\K_\mathrm{S}^0$ 
ratio shown in \figRef{fig:LHCpTspectra}, $\meanPT$ as a function 
of the hadron mass and the charged multiplicity, both inclusive and for 
different hadrons, shown in \figRef{fig:LHCpTvsMorNch}, and the ratio of 
yields with respect to $(\pi^++\pi^-)$ as a function of the charged 
multiplicity for different hadrons, shown in \figRef{fig:LHCyieldRatios}.
The predictions of default \textsc{Pythia} are compared to the Gaussian 
and thermodynamical model with the modifications outlined in 
\secRef{sec:newModels}. 

Default \textsc{Pythia} describes the ATLAS charged particle $\pT$ 
distribution very well for values of $\pT>1~\text{GeV}/c$, but shows a 
bump at around $0.5~\text{GeV}/c$. The Gaussian model with modifications 
gives a similar shape and reduces the bump somewhat, while undershooting 
the distribution large $\pT$ by a few $\%$. The thermodynamical model 
improves the description quite substantially, especially for low-$\pT$ 
values, where the aforementioned bump is almost gone. The predictions for 
the CMS charged hadron $\pT$ spectrum behave mostly similar, with the same 
bump visible for default \textsc{Pythia} and the Gaussian model.

For default \textsc{Pythia}, pions obtain a too hard $\pT$ spectrum.
The modifications to the Gaussian model improve the distribution
slightly, but there is still no good overall description. With the 
thermodynamical model the spectrum improves for low-$\pT$ values
quite a bit; however, it is still a bit too high in the large-$\pT$ 
region. The $\K^\pm$ $\pT$ spectrum shows the opposite behaviour: 
too many soft and too few hard kaons. The Gaussian model with 
modifications improves the description in the soft region somewhat, 
compared to default \textsc{Pythia}. Both the Gaussian and thermodynamical 
model change the shape of the spectrum slightly, but do not provide 
a better overall description of the $\K^\pm$ $\pT$.
The prediction of both models for the $\p,\bar\p$ $\pT$ spectrum are 
better compared to default \textsc{Pythia}, where 
especially the thermodynamical model improves the low-$\pT$ region
quite substantially.

While the prediction of the thermodynamical model for the $\Lambda/\K_
\mathrm{S}^0$ ratio is somewhat flatter with respect to the data, especially 
in the low-$\pT$ region, the normalization is off by almost a factor
of two due to the combination of producing slightly too many $\K_\mathrm{S}^0$ 
and not enough $\Lambda$. The observable could be improved by adjusting the 
overall normalization factor of baryons with respect to mesons. The value of 
this parameter has been fixed using the proton $\pT$ spectrum, however.

All models give very similar predictions for $\meanPT$ as a 
function of $n_\mathrm{ch}$, with an extremely good description of the 
region $n_\mathrm{ch}>20$, but too low $\meanPT$ for smaller 
multiplicities. It is quite obvious that the description of this observable
could be further improved by choosing a larger value for the width or 
temperature respectively and simultaneously lowering the $n_\text{string}
^\text{eff}$-dependence and hadron rescattering. This would however come 
hand in hand with worse descriptions of other observables.
For default \textsc{Pythia}, pions obtain a too large $\meanPT$ 
and heavier hadrons a too small one. While the thermodynamical model 
improves the predictions for $\langle\pT\rangle(m)$, there is still no
full agreement with data. The Gaussian model lies in between default 
\textsc{Pythia} and the thermodynamical model. We observe a similar 
behaviour for the $\langle\pT\rangle(n_\mathrm{ch})$ distribution 
for individual hadrons. The pion $\meanPT$ is described fairly
well with a slope that is slightly too steep. The main difference of 
the other hadrons with respect to pions is that they obtain a too small
$\meanPT$ over the whole $n_\mathrm{ch}$ range. As for pions, 
the slopes tend to be too steep.

ALICE~\cite{Adam:2016emw} found that the production of strange and 
multi-strange hadrons is enhanced with increasing multiplicity. While 
default \textsc{Pythia} is not able to reproduce such a behaviour, 
\figRef{fig:LHCyieldRatios} shows that the thermodynamical model achieves an 
increase of strangeness with charged multiplicity for $\K_\mathrm{S}^0$, 
$\Lambda$, and $\Xi$, but not for $\Omega$. Except for the latter, we
therefore expect the thermodynamical model to give an improved description 
of the data presented in \cite{Adam:2016emw}. The Gaussian model with 
modifications shows the opposite effect, a decrease with growing multiplicity.
These findings can be explained as follows: In default \textsc{Pythia} all 
(primary) hadrons are produced with a probability that is independent of the 
multiplicity or number of strings. In the thermodynamical model heavier 
hadrons are produced preferably at large $\pT$ values. Including the 
$n_\mrm{string}^\mrm{eff}$-dependence leads to potentially higher temperatures 
for events with large $n_\text{ch}$, where heavy hadrons have a higher 
probability to be produced, compared to low-$n_\text{ch}$ events. With the 
modifications to the Gaussian model, all hadrons obtain more $\pT$ in 
events with large values of $n_\mrm{string}^\mrm{eff}$. Due to phase-space 
constraints heavier hadrons might be rejected more often compared to 
lower-mass hadrons, leading to the decrease with growing multiplicity.
This might also be the reason for the slight drop towards large $n_
\text{ch}$ for $\Omega$ in the thermodynamical model, as it eventually
dominates over the effect of the $n_\mrm{string}^\mrm{eff}$-dependence.

\subsubsection{LEP and SLC}

While the main motivation for introducing the exponential $\pT$
distribution is arising from LHC data, the valid question of whether
the same model is able to describe $\e^+\e^-$ observables as well remains.
The effect of the close-packing of strings and hadron rescattering
are not included for $\e^+\e^-$ data as we do not expect them to represent
relevant physics here. Furthermore the string dependence relies on 
rapidity differences and an event axis aligned with the beam, which is
not present in $\e^+\e^-$ collisions. 

\begin{figure}[p!]
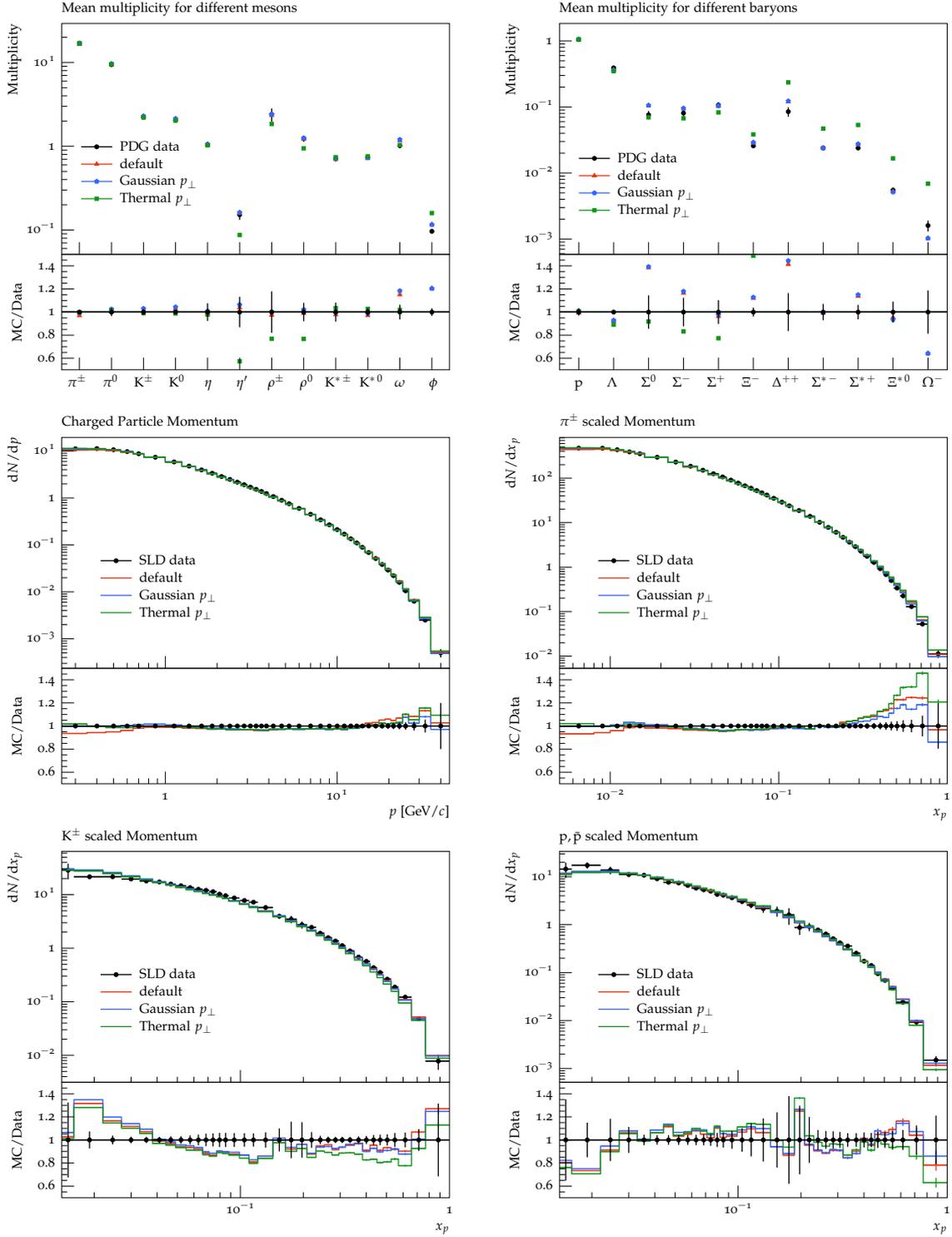

\centering
\rivetFigure{data/PDG/d03-x01-y03-meanMultiMesons.pdf} \hfill
\rivetFigure{data/PDG/d03-x01-y04-meanMultiBaryons.pdf} \\
\rivetFigure{data/SLD_2004_S5693039/d01-x01-y01-chPartX.pdf} \hfill
\rivetFigure{data/SLD_2004_S5693039/d02-x01-y02-piX.pdf} \\
\rivetFigure{data/SLD_2004_S5693039/d03-x01-y02-kX.pdf} \hfill
\rivetFigure{data/SLD_2004_S5693039/d04-x01-y02-pX.pdf} \\
\caption{Mean hadron multiplicities (\textit{top}), charged particle 
momentum (\textit{middle left}), and scaled momenta $x_p=2|\mathbf p|
/E_\mrm{cm}$ of $\pi^\pm$
(\textit{middle right}), $\K^\pm$ (\textit{bottom left}) and 
$\p,\bar\p$ (\textit{bottom right}).
Predictions of default \textsc{Pythia}~8, the Gaussian and 
thermodynamical model compared to PDG~\cite{Amsler:2008zzb} and 
SLD data~\cite{Abe:2003iy}.
\label{fig:LEPdata1}}
\end{figure}

\begin{figure}[t!]
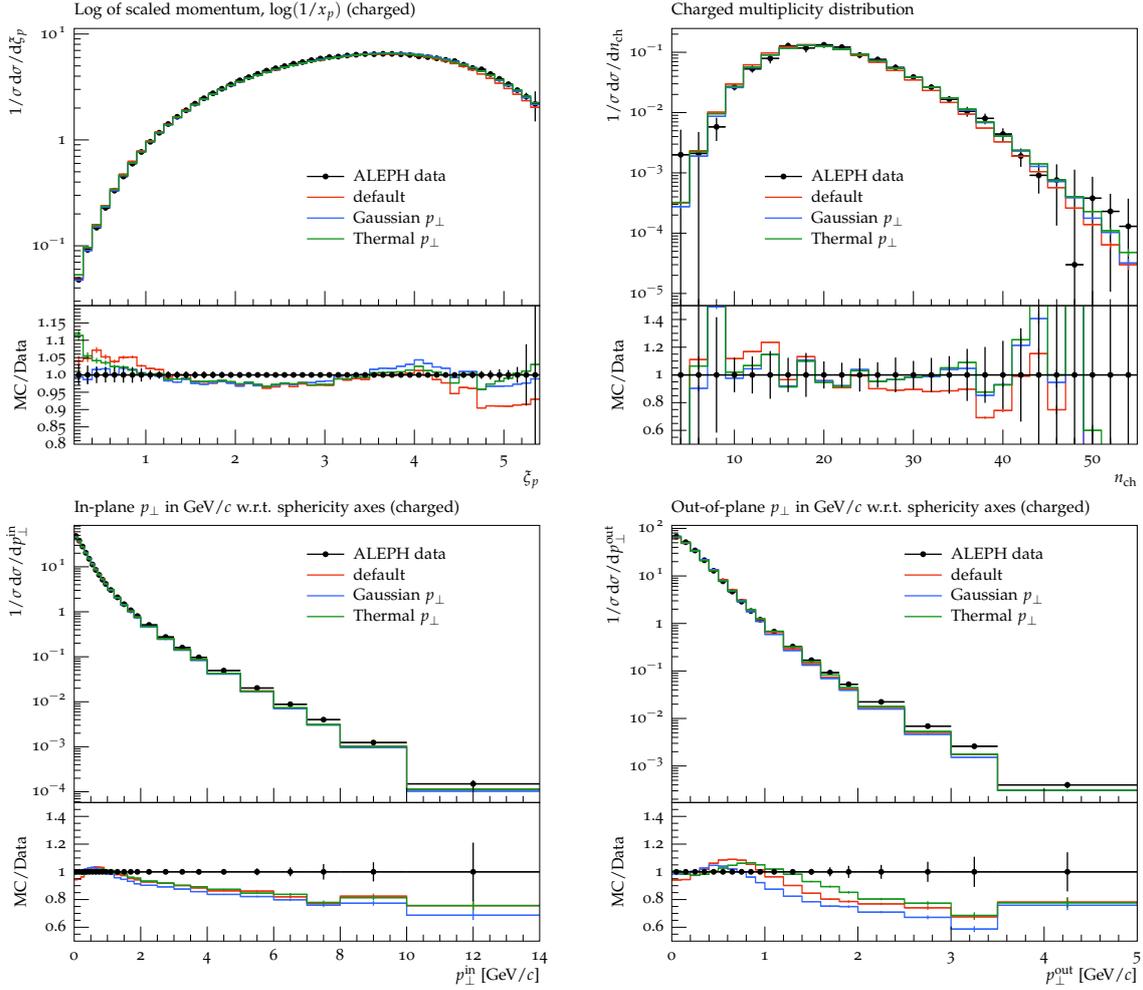

\centering
\rivetFigure{data/ALEPH_1996_S3486095/d17-x01-y01-logX.pdf} \hfill
\rivetFigure{data/ALEPH_1996_S3486095/d18-x01-y01-nCh.pdf} \\
\rivetFigure{data/ALEPH_1996_S3486095/d11-x01-y01-pTin.pdf} \hfill
\rivetFigure{data/ALEPH_1996_S3486095/d12-x01-y01-pTout.pdf} \\
\caption{Charged multiplicity distribution (\textit{top left}),
scaled momentum (\textit{top right}) and the inclusive
$p_{\perp\mathrm{in}}$ and $p_{\perp\mathrm{out}}$ spectra (\textit{bottom}).
Predictions of default \textsc{Pythia}~8, the Gaussian and 
thermodynamical model compared to ALEPH data~\cite{Barate:1996fi}.
\label{fig:LEPdata2}}
\end{figure}

The parameters of the thermodynamical model are adjusted using the charged
particle momentum spectrum as well as the scaled momenta of $\pi^\pm$, 
$\K^\pm$ and $\p,\bar\p$, measured with SLD~\cite{Abe:2003iy}, while the 
ALEPH event shapes~\cite{Barate:1996fi} served as cross checks. For the 
Gaussian model the width and prefactors for strange and diquarks have been
adjusted such that the mean charged multiplicities agrees with the value
obtained with the thermodynamical model. The corresponding settings and 
values can be found in \appRef{app:settings}. The $\e^+\e^-$ data set 
presented here includes the aforementioned momenta and mean multiplicities 
for different hadrons shown in \figRef{fig:LEPdata1}, as well as the 
charged multiplicity distribution, scaled momentum and the inclusive 
$p_{\perp\mathrm{in}}$ and $p_{\perp\mathrm{out}}$ spectra shown in 
\figRef{fig:LEPdata2}. The predictions of default \textsc{Pythia} are 
compared to the Gaussian and thermodynamical model as described above.

The only difference between default \textsc{Pythia} and the prediction 
labelled as ``Gaussian $\pT$'' is an adjusted value for the Gaussian
width and its prefactors for $\s$ and diquarks, i.e. there is no 
change of the flavour selection 
parameters. Therefore, the values of the mean multiplicities remain, 
leading to overlapping data and Monte Carlo histogram points in 
\figRef{fig:LEPdata1}, which are thus not fully visible. With the 
thermodynamical model we obtain a fairly good description of most 
hadrons, with the notable exceptions of producing too many heavy
baryons.
Note however, that the Gaussian model comes with around 20 parameter for 
selecting the flavour of new hadrons, whereas the thermodynamical model makes 
use of only three parameters: the temperature, the overall normalization 
factor of baryons with respect to mesons, and the additional suppression 
factor for hadrons with strange quarks, see \secRef{sec:thermModel}. Hence,
the result is fairly acceptable.

The predictions of the two models for the charged particle momentum 
agree very well with data in the soft region; there is only some small 
deviation for medium and large momenta, where especially the 
thermodynamical model predicts somewhat too many particles in the hard 
region. The same effect is even clearer visible in the scaled momentum 
spectrum of pions. 
For kaons and protons we observe the opposite effect: the new model
predicts too few hadrons with large momenta.

Similar to the charged particle momentum, the predictions of both models
for the logarithm of the scaled momentum agrees well with data, with
some small deviation for medium and large values. However, the sudden 
drop in the ratio of the Monte Carlo prediction to data at around 
$\xi_p=4.7$ remains almost unchanged.
The description of the charged multiplicity distribution improves
slightly, compared to default \textsc{Pythia}, towards having less 
events with small multiplicities and more events with larger ones.
This is due to having an increased mean charged multiplicity.
While both models describe the low-$\pT$ region of the inclusive 
$p_{\perp\mathrm{in}}$ and $p_{\perp\mathrm{out}}$ spectra very 
well, they underestimate the amount of events with larger $\pT$
values. The thermodynamical model provides a better description
of especially the $p_{\perp\mathrm{out}}$ spectrum, compared to 
the Gaussian model.

To summarize we note that the thermodynamical model is able to 
provide predictions for event shapes and momentum spectra in 
$\e^+\e^-$ events that are of a similar quality as those by the 
Gaussian model. Nevertheless, the hadron decomposition is not 
described well, a price to pay for reducing the amount of 
flavour selection parameters.

\section{Summary and Outlook}
\label{sec:summary}

The understanding of soft hadronic physics is changing under the
onslaught of LHC pp data. Of course, there has never been an
approach that could describe all aspects of pp physics perfectly,
but before LHC it was often assumed that all the basic concepts were
in place, and that what remained was successive refinements.
Now we see that there is still much left to learn. There have already
been several surprises, and further data analyses may well produce
more.

In view of this we have revisited some of the basic soft-physics
assumptions of the \textsc{Pythia} event generator, which has
been quite successful in predicting and describing many aspects
of the data, but now starts to show cracks. New approaches have
here been studied for some areas, to understand how much room for
improvements there would be, without any claim that either of
them would necessarily be the one and only right way to go.

A central pillar of \textsc{Pythia} has been the Lund string
fragmentation model, where a tunneling mechanism for string breakups
leads to a universal Gaussian $\pT$ spectrum. In this work a
thermodynamical model is implemented as an alternative, where
$\pT$ instead follows an exponential distribution. For an
already selected $\pT$, the hadron flavour is picked based
on an exponential $m_{\perp}$ weight, with additional factors
due to spin-counting rules and so on. This approach suppresses
the production of heavier hadrons, and gives them a larger
$\meanPT$. Such a pattern is observed in data,
and exists in the Gaussian approach mainly owing to particle decays,
but there undershoots data.

Making the Gaussian $\pT$ width, or temperature in case of the
thermodynamical model, dependent on the close-packing of strings
allows for modelling the influence of strings on each other in a simple
way. An effective number of density of strings is introduced for low
$\pT$'s, while high-$\pT$ fragmentation tends to occur
outside the close-packed string region and is left unaffected. Such a
mechanism could e.g. be used to explain a changing flavour composition
at high multiplicities.

Finally we implemented a simple model for hadronic rescattering,
applied to the primary hadrons, before decays. The probability of two
hadrons to rescatter is based on how close they are in phase space.
By favouring a shift towards equal transverse velocities, it should
also give higher $\meanPT$ for heavy hadrons and
lower for pions.

Not surprisingly we found the hadronic decays to limit the hoped-for
effects. Specifically, most pions come from decays of heavier hadrons,
and so the mechanisms intended to give less $\pT$ to pions and
more to kaons and protons are largely nullified. The mechanisms are
also not simply additive; starting out from the thermodynamical model
with its already-existing mass differentiation, the further effects
of varying temperatures or hadronic rescattering are smaller than
corresponding effects in the Gaussian approach.

Nevertheless the thermodynamical model is able to provide reasonable
descriptions of observables such as the $\pT$ spectrum of charged
hadrons, the average transverse momentum as a function of the hadron 
mass, or the recently measured enhanced production of strange and 
multi-strange hadrons with increasing multiplicity. These observables 
have so far been described rather poorly by \textsc{Pythia}. And, given 
the small number of flavour parameters in the thermodynamical model, 
it is able to describe a reasonable number of $\ee$ data rather well, 
even if it can not compete with the many-more-parameter tunes of 
default \textsc{Pythia}.

It should be noted that we have not compared with all relevant
available data, by far. Notably, the ridge effect was not described
by the existing \textsc{Pythia} model, and our current changes do
not introduce any mechanism to induce it. The ridge was first
observed in AA collisions
\cite{Adams:2005ph,Abelev:2009af,Alver:2009id,Chatrchyan:2011eka},
where nuclear geometry and hydrodynamical expansion offer natural
starting points \cite{Ollitrault:1992bk,Voloshin:1994mz}, although
the range of detailed models is too vast to cover here
\cite{Akiba:2015jwa}. In the field of pp physics \cite{Li:2012hc},
the EPOS model addresses the issue by having an inner core that can
push strings in the outer corona \cite{Werner:2010ss}, whereas a
recent extension of DIPSY \cite{Bierlich:2016vgw} provides a
corresponding shove from the excess energy of central overlapping
strings that form ropes. In a similar spirit, our higher string
tension could introduce a push also without rope formation.
A detailed modelling is not trivial, however, and we have not
pursued it for now.

To advance to the next level of sophistication within the line of
research advocated here, it would be necessary to do a microscopic
tracing of the full space--time evolution of the event, both for
partons and for hadrons, and including both production and decay
vertices. This is nontrivial beyond the simple one-string picture,
even in the cleaner $\ee$ events, and the further complications
of MPIs and CR in hadronic events will make it even worse. What it
would allow is a more detailed understanding of the close-packing
both of strings and of hadrons. Combined with a more detailed modelling
of hadronic rescattering, a more realistic picture may emerge.

Some of the limitations encountered here are likely still to remain,
so further mechanisms may be at play, in addition to the ones studied
here. This would not be the first time where a cocktail of smaller
effects combine to give a significant signal. What is less likely is
actually the opposite, that one single mechanism does it all.
Specifically, whatever else may be going on, the close-packing of
strings and hadrons appears unavoidable in high-multiplicity pp events,
and collective-flow effects are here to stay. In sum, we have an
interesting and challenging time ahead of us, where some of the most
unexpected new LHC observations may well come in the low-$\pT$
region rather than the in high-$\pT$ one.

\section*{Acknowledgements}

Helpful discussions with Peter Skands are gratefully acknowledged.
Work supported by the MCnetITN FP7 Marie Curie Initial Training Network, 
contract PITN-GA-2012-315877. This project has also received funding 
in part by the Swedish Research Council, contract number 621-2013-4287, 
and in part from the European Research Council (ERC) under the European 
Union's Horizon 2020 research and innovation programme (grant agreement 
No 668679).

\appendix

\section{Settings} \label{app:settings}

\TabRef{tab:settings} gives a list of the settings that have been 
changed with respect to default \textsc{Pythia} to obtain the 
results presented in \secRef{sec:finalResults}.

\begin{table}[t!]
\caption{\textsc{Pythia}~8 parameters and their values for tuning
to LHC and LEP/SLC observables.
\label{tab:settings}}
\begin{flushleft}
\begin{tabular}{llllll}
\toprule
& & \multicolumn{2}{c}{\textbf{LHC}} &
\multicolumn{2}{c}{\textbf{LEP and SLC}} \\
& \small{Default} & \small{Gaussian} & \small{Thermal} & 
\small{Gaussian} & \small{Thermal} \\
\midrule
Switch to thermal model? \\
\texttt{StringPT:thermalModel} & 
\texttt{off} & \texttt{off} & \texttt{on} & \texttt{off} & \texttt{on} \\[1mm]
Gaussian width $\sigma=\sqrt{\kappa/\pi}$ \\
\texttt{StringPT:sigma} & 
\texttt{0.335} & \texttt{0.33} & \texttt{-} & \texttt{0.295} & 
\texttt{-} \\[1mm]
$\sigma$ prefactor for $\s$ quarks \\
\texttt{StringPT:widthPreStrange} & 
\texttt{1.0} & \texttt{1.2} & \texttt{-} & \texttt{1.2} & \texttt{-} \\[1mm]
$\sigma$ prefactor for diquarks \\
\texttt{StringPT:widthPreDiquark} & 
\texttt{1.0} & \texttt{1.2} & \texttt{-} & \texttt{1.2} & \texttt{-} \\[1mm]
Fraction with enhanced $\sigma$ \\
\texttt{StringPT:enhancedFraction} & 
\texttt{0.01} & \texttt{0.0} & \texttt{-} & \texttt{0.0} & \texttt{-} \\[1mm]
Temperature $T$ \\
\texttt{StringPT:temperature} & 
\texttt{-} & \texttt{-} & \texttt{0.21} & \texttt{-} & \texttt{0.205} \\[1mm]
Baryon normalization factor \\
\texttt{StringFlav:BtoMratio} & 
\texttt{-} & \texttt{-} & \texttt{0.357} & \texttt{-} & \texttt{0.625} \\[1mm]
Suppression factor for $\s$ hadrons \\
\texttt{StringFlav:StrangeSuppression} & 
\texttt{-} & \texttt{-} & \texttt{0.5} & \texttt{-} & \texttt{0.45} \\[1mm]
$r$ parameter in \eqRef{eq:nStringEffKappa} or \eqref{eq:nStringEffTemp} \\
\texttt{StringPT:expNSP} & 
\texttt{0.0} & \texttt{0.01} & \texttt{0.13} & \texttt{0.0} & 
\texttt{0.0} \\[1mm]
Range of MPI-based CR scheme \\
\texttt{ColourReconnection:range} & 
\texttt{1.8} & \texttt{1.8} & \texttt{1.1} & \texttt{-} & \texttt{-} \\[1mm]
Hadron rescattering (HR) on? \\
\texttt{HadronLevel:HadronScatter} & 
\texttt{off} & \texttt{on} & \texttt{on} & \texttt{off} & \texttt{off} \\[1mm]
HR with \eqRef{eq:mCut} and \eqref{eq:scatProbDS} \\
\texttt{HadronScatter:mode} & 
\texttt{-} & \texttt{0} & \texttt{0} & \texttt{-} & \texttt{-} \\[1mm]
$P^\mathrm{max}_{\,\mathrm{ds}}$ parameter in \eqRef{eq:scatProbDS} \\
\texttt{HadronScatter:maxProbDS} & 
\texttt{-} & \texttt{0.25} & \texttt{0.5} & \texttt{-} & \texttt{-} \\[1mm]
MPI regularization parameter \\
\texttt{MultipartonInteractions:pT0Ref} &
\texttt{2.28} & \texttt{2.34} & \texttt{2.5} & \texttt{-} & \texttt{-} \\
\bottomrule
\end{tabular}
\end{flushleft}
\end{table}

\bibliographystyle{JHEP}
\bibliography{tf.bbl}

\end{document}